\documentclass{osa-article}

\journal{osac}

\articletype{Research Article}

\usepackage{lineno}
\begin{document}
\title{Hexapartite steering based on a four-wave-mixing process with a spatially structured pump}

\author{Yunyun Liang,\authormark{1,2}, Rongguo Yang\authormark{1,2,3*}, Jing Zhang,\authormark{1,2,3*}, and Tiancai Zhang,\authormark{1,3}}

\address{\authormark{1}State Key Laboratory of Quantum Optics and Quantum Optics Devices, Shanxi University, Taiyuan
030006, China\\
\authormark{2}College of Physics and Electronic Engineering, Shanxi University, Taiyuan 030006, China\\
\authormark{3}Collaborative Innovation Center of Extreme Optics, Shanxi University, Taiyuan, 030006, China}

\email{\authormark{*}zjj@sxu.edu.cn{*}
yrg@sxu.edu.cn} 



\begin{abstract}
Multipartite Einstein-Podolsky-Rosen (EPR) steering has been widely studied, for realizing safer quantum communication. The steering properties of six spatially separated beams from the four-wave-mixing process with a spatially structured pump are investigated. Behaviors of all (1+i)/(i+1)-mode (i=1,2,3) steerings are understandable, if the role of the corresponding relative interaction strengths are taken into account. Moreover, stronger collective multipartite steerings including five modes also can be obtained in our scheme, which has potential applications in ultra-secure multiuser quantum networks when the issue of trust is critical. By further discussing about all monogamy relations, it is noticed that the type-IV monogamy relations, which are naturally included in our model, are conditionally satisfied. Matrix representation is used to express the steerings for the first time, which is very useful to understand the monogomy relations intuitively. Different steering properties obtained in this compact phase-insensitive scheme have potential applications for different kinds of quantum communication tasks. 
\end{abstract}

\section{Introduction}

Quantum correlation is one of the central concepts in quantum information theory \cite{Zeilinger1993}, and there is often a clear hierarchy among diﬀerent correlations, which includes total correlation, discord, entanglement, steering, and Bell nonlocality \cite{Guo2020}. Einstein-Podolsky-Rosen (EPR) steering\cite{Uola2020}, which was ﬁrst proposed by Schrödinger in 1935 \cite{Schrodinger1935} and developed with concrete definition and criterion by Wiseman in 2007 \cite{Wiseman2007,Wiseman2009}, denotes a quantum correlation situated between entanglement and Bell nonlocality, i.e., EPR steering is stricter than quantum entanglement. It is identified as a significant physical resource for one-sided device-independent (1SDI) quantum key
distribution \cite{Wiseman2012,Wiseman2016}, secure quantum teleportation \cite{reid2013,he2015secure}, and subchannel discrimination \cite{piani2015}, due to its natural asymmetry, which means that Alice’s ability to steer Bob may not be equal to Bob’s ability to steer Alice. The situation that only one part can steer the other is called one-way steering\cite{Olsen2010,Roman2012}, which has potential applications in hierarchical quantum communication. The situation that both parts can steer each other is called two-way steering, which was proved to achieve quantum teleportation with fidelity greater than 0.67\cite{He2015}. Besides bipartite steering, multipartite steering was also defined\cite{Wisema2011} and the corresponding criterion was given theoretically \cite{He2013mult}  and then verified experimentally\cite{He2015mult}. Multipartite steering, which is important for scalable quantum network, can be generated in many physical systems, such as cavity optomechanical system \cite{Li2017,Tan2021,He2013-1}, optical parametric process  \cite{Su2017,He2020fc}, and four-wave mixing (FWM) system\cite{Jing2017,He2020mo}. The four-wave mixing (FWM) process can be used to generate squeezed\cite{Alberto2019} or entangled state of light\cite{Lett2008,Gao2018,Jing2014}, and can be further generate multipartite entanglement when cascaded FWM is applied\cite{Jing2014,Jing2018-1}. Genuine tripartite steering, one-way steering, and collective steering based on asymmetric cascaded FWM structure were discussed \cite{Jing2017,He2019}. Quadripartite steering generated from the symmetric and asymmetric cascaded FWM structures was demonstrated and four distinct types of monogamy relations were analyzed accordingly\cite{He2020mo}. By adding one more FWM, pentapartite steering can be obtained and collective multipartite steering can only exist in the asymmetry structure when optical loss is considered\cite{Yin2021}. To achieve better scalability, FWM process with a spatially structured pump (SSP)\cite{Fabre2020,Jing2022}, which can be realized by altering the angle between two or more pump beams, was adopted to generate quantum correlations among six\cite{Jing2017} , ten\cite{Jing2018-2}, fourteen\cite{Jing2019} output beams experimentally. Besides scalability, using spatially structured pump can make the experimental setup simpler and compacter. In this paper, we investigate the steering ability of six spatially separated output beams from a FWM process with a spatially structured pump\cite{Jing2020-1}, and demonstrate the corresponding (1+i)/(i+1)-mode (i=1,2,3) steerings, collective steerings, and monogamy relations.

\section{Physical Model and Theoretical Derivation}

\begin{figure*}[htbp]
	\begin{minipage}{0.32\linewidth}
		\vspace{3pt}
		\centerline{\includegraphics[height=4.2cm,width=4.5cm]{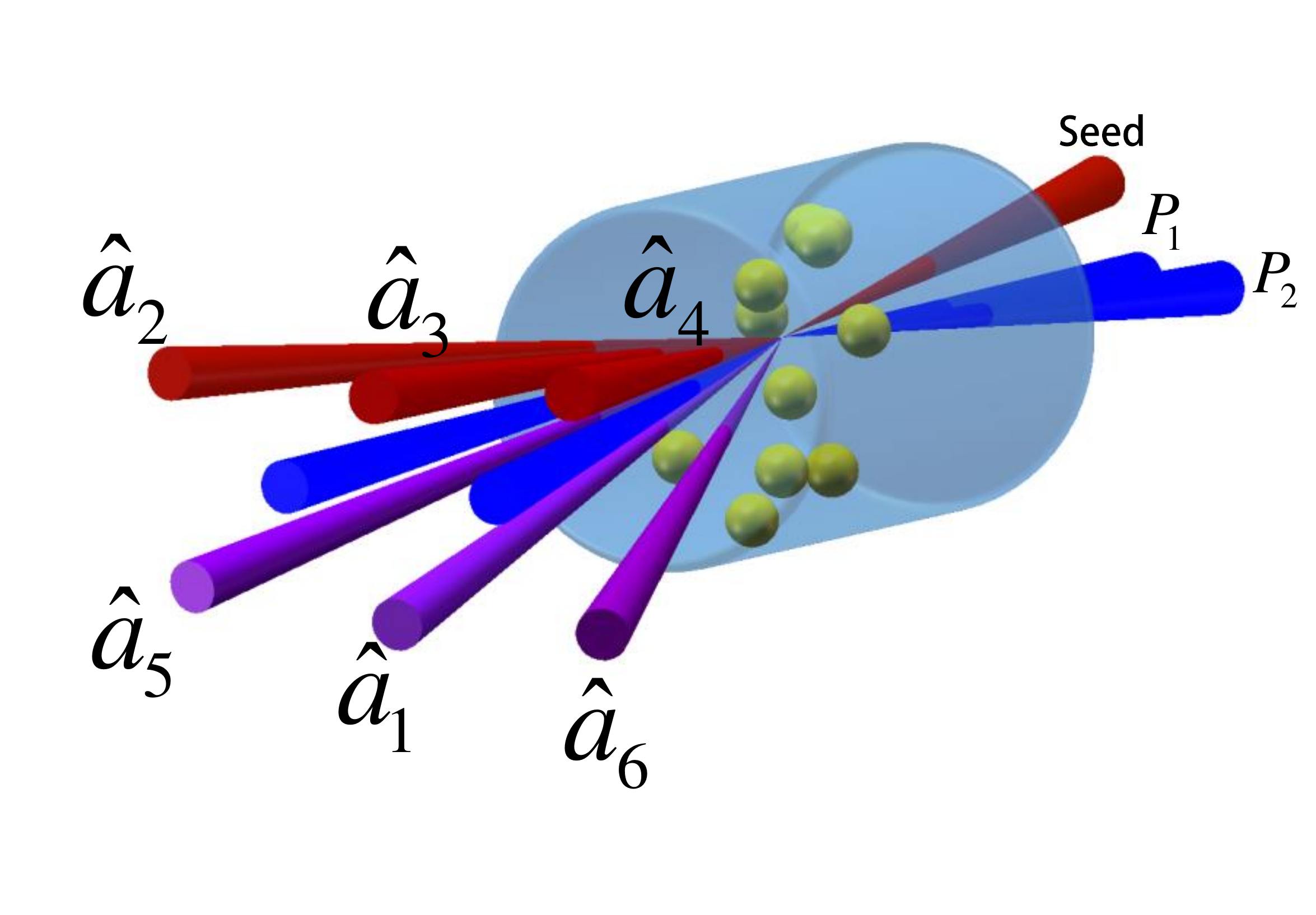}}
	\end{minipage}
	\put(-60,65){\bf(a)}
\put(55,65){\bf(b)}
\put(175,65){\bf(c)}
	\begin{minipage}{0.32\linewidth}
		\vspace{3pt} 
		\centerline{\includegraphics[height=4.0cm,width=4.5cm]{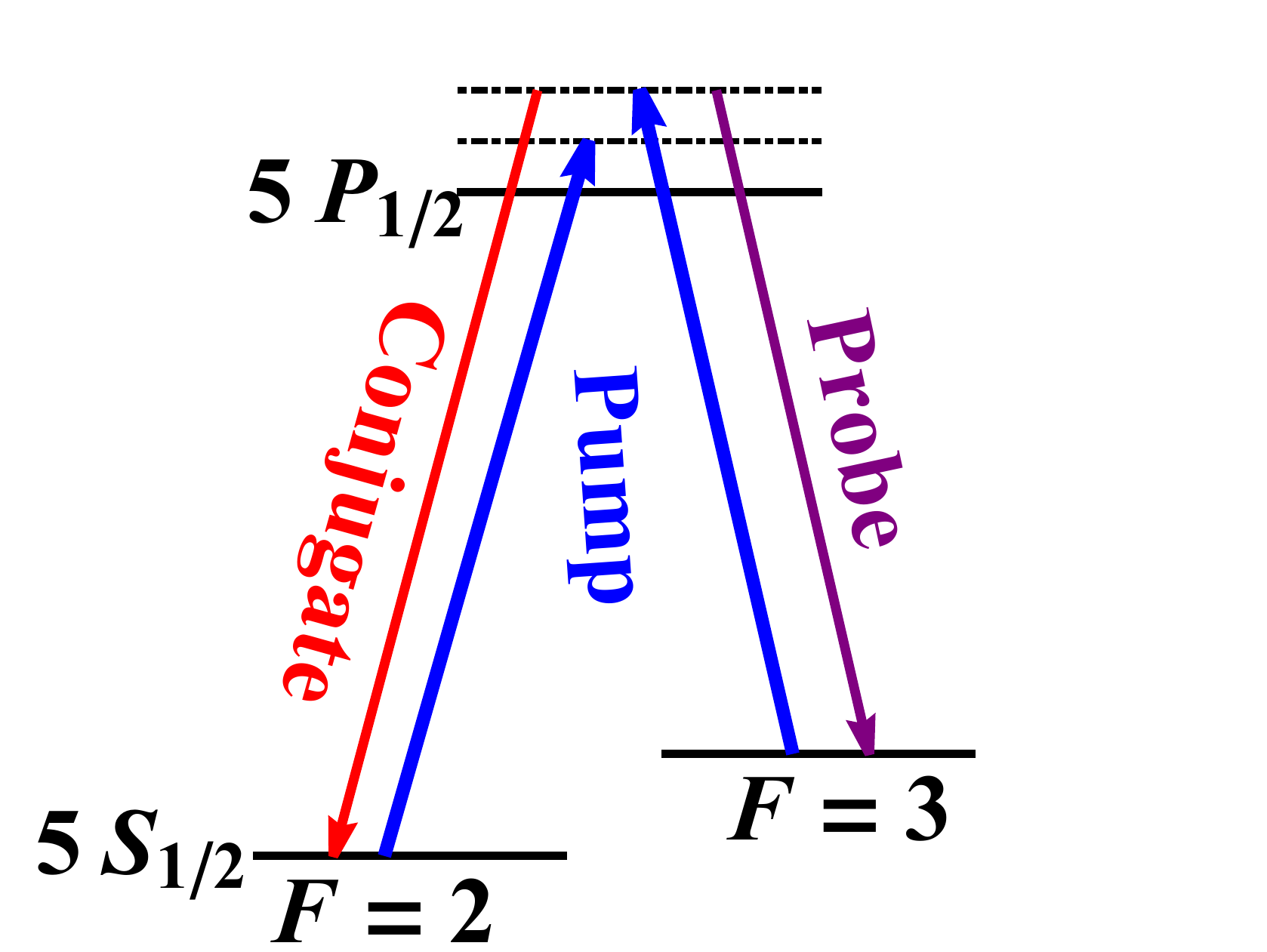}}
       
	\end{minipage}
	\begin{minipage}{0.32\linewidth}
		\vspace{3pt}
		\centerline{\includegraphics[height=4.3cm,width=4.3cm]{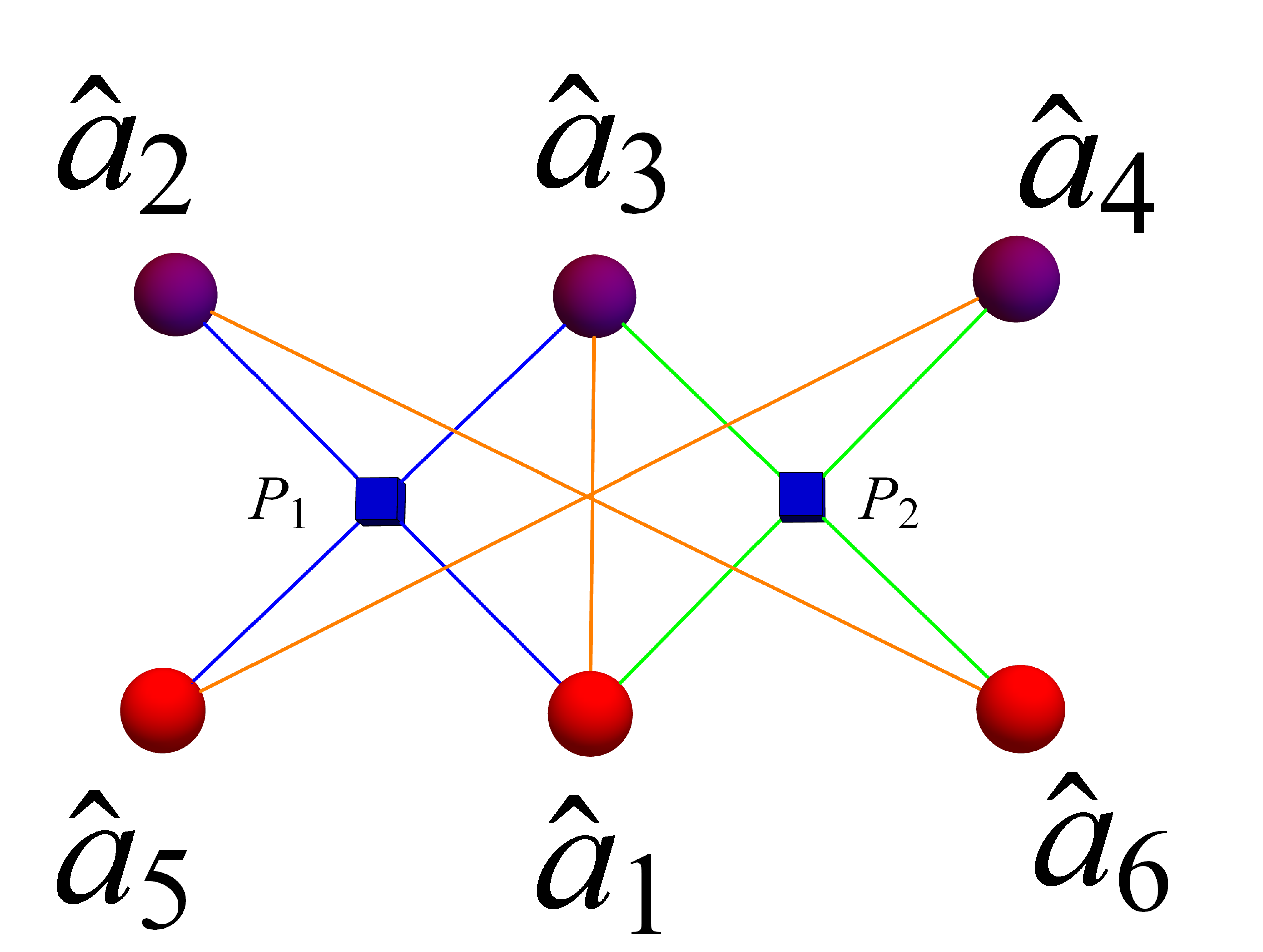}}
	\end{minipage}
	
\caption{The FWM process with a spatially structured pump. (a) Generated six beams after FWM process with structured pump1 and pump2, (${\hat{a}_{2}},{\hat{a}_{3}},{\hat{a}_{4}}$) and (${\hat{a}_{1}},{\hat{a}_{5}},{\hat{a}_{6}}$)  are the probe and conjugate beams, respectively. (b) The corresponding energy-level diagram of the double-$\Lambda $ scheme in the D1 line of ${}^{85}Rb$ vapor cell. (c) The  output beams from the SSP based on FWM process. The red and purple balls represent the probe and conjugate beams, separately. The intersections of the blue, green, and orange lines correspond to pump1, pump2, and double pump, separately.}       
\end{figure*}
We consider a FWM process with a spatially structured pump, as is shown in Fig.1(a). Two bright pump beams (pump1 and pump2) and a weak seed beam are focused in the center of the ${}^{85}Rb$  vapor cell with a small tilted angle. Three probe  beams (${\hat{a}_{2}},{\hat{a}_{3}},{\hat{a}_{4}}$) and three conjugate beams  (${\hat{a}_{1}},{\hat{a}_{5}},{\hat{a}_{6}}$)   are simultaneously generated and naturally separated in space. The corresponding atomic energy level diagram is shown in Fig.1(b). According to the phase-matching conditions, single-pump and double-pump FWM processes both happen. Therefore, correlations between ${\hat{a}_{1}} $and $ {\hat{a}_{2}}$, ${\hat{a}_{3}} $and $ {\hat{a}_{5}}$, ${\hat{a}_{1}} $and $ {\hat{a}_{4}}$, ${\hat{a}_{3}} $and $ {\hat{a}_{6}}$, are generated through single-pumped processes, and correlations between  ${\hat{a}_{1}} $and $ {\hat{a}_{3}}$, ${\hat{a}_{4}} $and $ {\hat{a}_{5}}$, ${\hat{a}_{2}} $and $ {\hat{a}_{6}}$, are generated through double-pumped processes,which can be seen in Fig.1(c). The interaction Hamiltonian can be written as
\begin{gather}
     \hat{H}=i\hbar [{{\varepsilon }_{1}}\hat{a}_{1}^{\dagger }\hat{a}_{2}^{\dagger }+{{\varepsilon }_{2}}\hat{a}_{1}^{\dagger }\hat{a}_{3}^{\dagger }+{{\varepsilon   }_{3}}\hat{a}_{1}^{\dagger }\hat{a}_{4}^{\dagger }+{{\varepsilon }_{4}}\hat{a}_{3}^{\dagger }\hat{a}_{5}^{\dagger }\notag\\
     +{{\varepsilon }_{5}}\hat{a}_{3}^{\dagger }\hat{a}_{6}^{\dagger }+{{\varepsilon }_{6}}\hat{a}_{4}^{\dagger }\hat{a}_{5}^{\dagger }+{{\varepsilon }_{7}}\hat{a}_{2}^{\dagger }\hat{a}_{6}^{\dagger }]+H.c.
\end{gather}
where ${\hat{a}^{\dagger}_{1}},...,{\hat{a}^{\dagger}_{6}}$ are the creation operators of six output beams. The first and forth terms represents the FWM processes of generating ${\hat{a}_{1}}$ and ${\hat{a}_{2}}$, ${\hat{a}_{3}}$ and ${\hat{a}_{5}}$ with a single pump (pump1). The third and fifth terms represents the generation of ${\hat{a}_{1}}$ and ${\hat{a}_{4}}$, ${\hat{a}_{3}}$ and ${\hat{a}_{6}}$ through single-pump (pump2) FWM processes. The second, sixth, and seventh terms corresponds to the double-pump (pump1 and pump2) FWM processes of generating ${\hat{a}_{1}}$ and $ {\hat{a}_{3}}$, ${\hat{a}_{4}}$ and $ {\hat{a}_{5}}$,  ${\hat{a}_{2}}$ and $ {\hat{a}_{6}}$ beams. $\varepsilon _{1}...\varepsilon _{7}$ represents the interaction strengths of seven different FWM processes. H.C. means the Hermitian Conjugate. Considering the symmetry of the output beams, we assume that ${{\varepsilon }_{1}}={{\varepsilon }_{4}}=G_1$, ${{\varepsilon }_{3}}={{\varepsilon }_{5}}=G_2, {{\varepsilon}_2}={{\varepsilon }_{6}}={{\varepsilon }_{7}}=G_3$, to make it convenient to calculate and understand. The Heisenberg equations governing the time evolution of the six beams can be written as
\begin{gather}
\frac{d\xi }{dt}=A\xi,
\end{gather}
where $\xi ={{({\hat{X}_{1}},{\hat{Y}_{1}}, ..., {\hat{X}_{6}},{\hat{Y}_{6}})}^T},$ here the definitions of the quadrature amplitude and phase operators are ${\hat{X}_{i}}=({{{\hat{a}}}_{i}}+\hat{a}_{i}^{\dagger } )$ and  ${\hat{Y}_{i}}=i( \hat{a}_{i}^{\dagger }-{{{\hat{a}}}_{i}} )$. The coefficient matrix $A$ is

\begin{gather}
A=\left(
\begin{array}{cccccc}
 0 & {{\varepsilon }_{1}} & {{\varepsilon }_{2}} & {{\varepsilon }_{3}} & 0 & 0 \\
 {{\varepsilon }_{1}} & 0 & 0 & 0 & 0 & {{\varepsilon }_{2}} \\
 {{\varepsilon }_{2}} & 0 & 0 & 0 & {{\varepsilon }_{1}} &{{\varepsilon }_{3}} \\
{{\varepsilon }_{1}} & 0 & 0 & 0 & {{\varepsilon }_{2}} & 0 \\
 0 & 0 & {{\varepsilon }_{1}} & {{\varepsilon }_{2}} & 0 & 0 \\
 0 & {{\varepsilon }_{2}} & {{\varepsilon }_{3}} & 0 & 0 & 0 \\
\end{array}
\right),
\end{gather}
 and satisfy that $A=P\Lambda {{P}^{-1}}$, whose diagonalized matrix is $\Lambda$ and the corresponding unitary matrix is $P$, yielding the solution $\xi =S\xi (0)$. $S$ can be obtained by
 $S=P{{e}^{\Lambda t }}{{P}^{-1}}$. Then the covariance matrix (CM) of the output six modes can be given by\cite{Loock2011}

\begin{gather}
\sigma =\left\langle \xi {{\xi }^{T}} \right\rangle =S\xi (0)\xi {{(0)}^{T}}{{S}^{T}}=S{{S}^{T}}
\end{gather}
Based on the CM, the EPR steering properties between the output beams can be quantified by steering criterion. For a bipartite Gaussian state system ( subsystem A contains $ n_A$ modes and subsystem B contains $n_B$ modes), the corresponding CM can be reconstructed in the form $\sigma _{AB}=\left(\begin{matrix}
   \mathcal{A} & \mathcal{C}  \\
{\mathcal{C}^T} & \mathcal{B}  \\
\end{matrix} \right)$, where submatrices $\mathcal{A}$ and $\mathcal{B}$ are the reduced state of subsystem $A$ and $B$, respectively, and submatrix $\mathcal{C}$ corresponds to the correlation between them. The steerability from subsystem A to subsystem B ($A \to B$) and from subsystem B to subsystem A ($B \to A$ ) can be defined as\cite{kogias2015quantification} 
\begin{equation}
    {\mathcal{G} ^{A\to B}}(\sigma _{AB})=\max \left\{ \left. 0,-\sum\limits_{j:\bar{\nu }_{j}^{\mathcal{AB/A}}<1}{\ln (\bar{\nu }_{j}^{\mathcal{AB/A}})} \right\} \right.
\end{equation}
\begin{equation}
{\mathcal{G}^{B\to A}}(\sigma _{BA})=\max \left\{ \left. 0,-\sum\limits_{j:\bar{\nu }_{j}^{\mathcal{AB/B}}<1}{\ln (\bar{\nu }_{j}^{\mathcal{AB/B}})} \right\} \right..
\end{equation}
where $\bar{\nu }_{j}^{\mathcal{AB/A}}(j=1,...,n_B)$ and $\bar{\nu }_{j}^{\mathcal{AB/B}}(j=1,...,n_A)$ are the symplectic eigenvalues of $\bar{\sigma} _{\mathcal{AB/A}}=\mathcal{B}-\mathcal{C}^T\mathcal{A}^{-1}\mathcal{C}$ and $\bar{\sigma} _{\mathcal{AB/B}}=\mathcal{A}-\mathcal{C}^T\mathcal{B}^{-1}\mathcal{C}$, respectively. B can be steered by A if $\mathcal{G} ^{A\to B}>0$. A can be steered by B if $\mathcal{G} ^{B\to A}>0$. When $({{\mathcal{G}}^{A\to B}}>0)$ and $({{\mathcal{G}}^{B\to  A}}>0)$ are both satisfied, the steering is two-way ( A and B can steer each other), otherwise $({{\mathcal{G}}^{A\to B}}>0,{{\mathcal{G}}^{B\to A}}=0)$ or $({{\mathcal{G}}^{B\to A}}>0,{{\mathcal{G}}^{A\to B}}=0)$ represents a one-way steering (only A can steer B, or only B can steer A).

\section{The (1+i)/(i+1)-mode EPR steering}
\subsection{The (1+1)-mode EPR steering}

Fig.2(a) gives the time evolution of all (1+1)-mode steerings and shows that $t=0.3$ can make all these steerings considerable. The (1+1)-mode steerings versus the interaction strength $G_1$ under different parameters are shown in Fig.2(b) ($G_2=1.2, G_3=2, t=0.3$) and  Fig.2(c) ($G_2=2, G_3=1.2, t=0.3$), respectively. 

\begin{figure}[htbp]
\centering
\quad
\put(60,96){\bf(a)}
\put(180,96){\bf(b)}
 \put(300,96){\bf(c)}
\hspace{0.2cm}
\includegraphics[height=3.2cm,width=4cm]{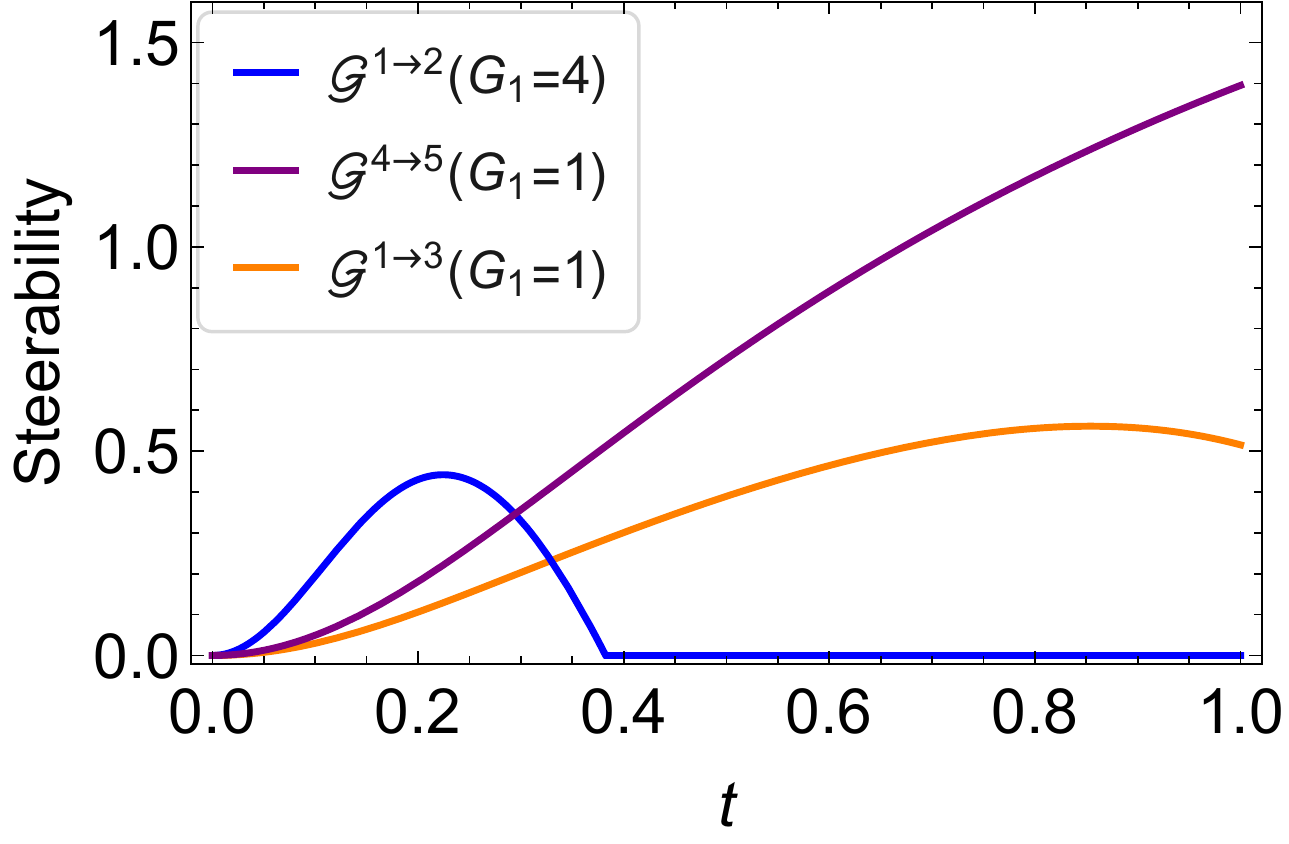}
\includegraphics[height=3.3cm,width=4cm]{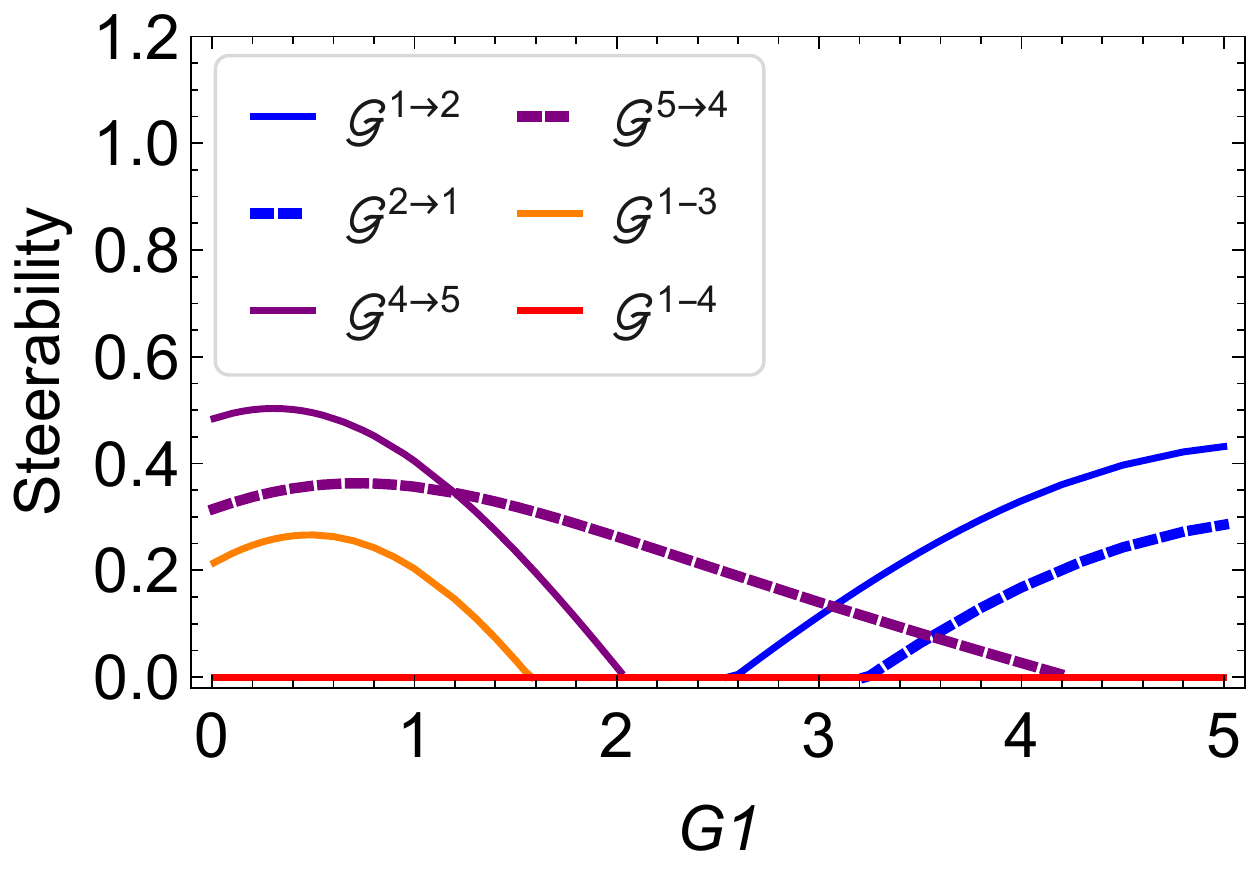}
\includegraphics[height=3.3cm,width=4cm]{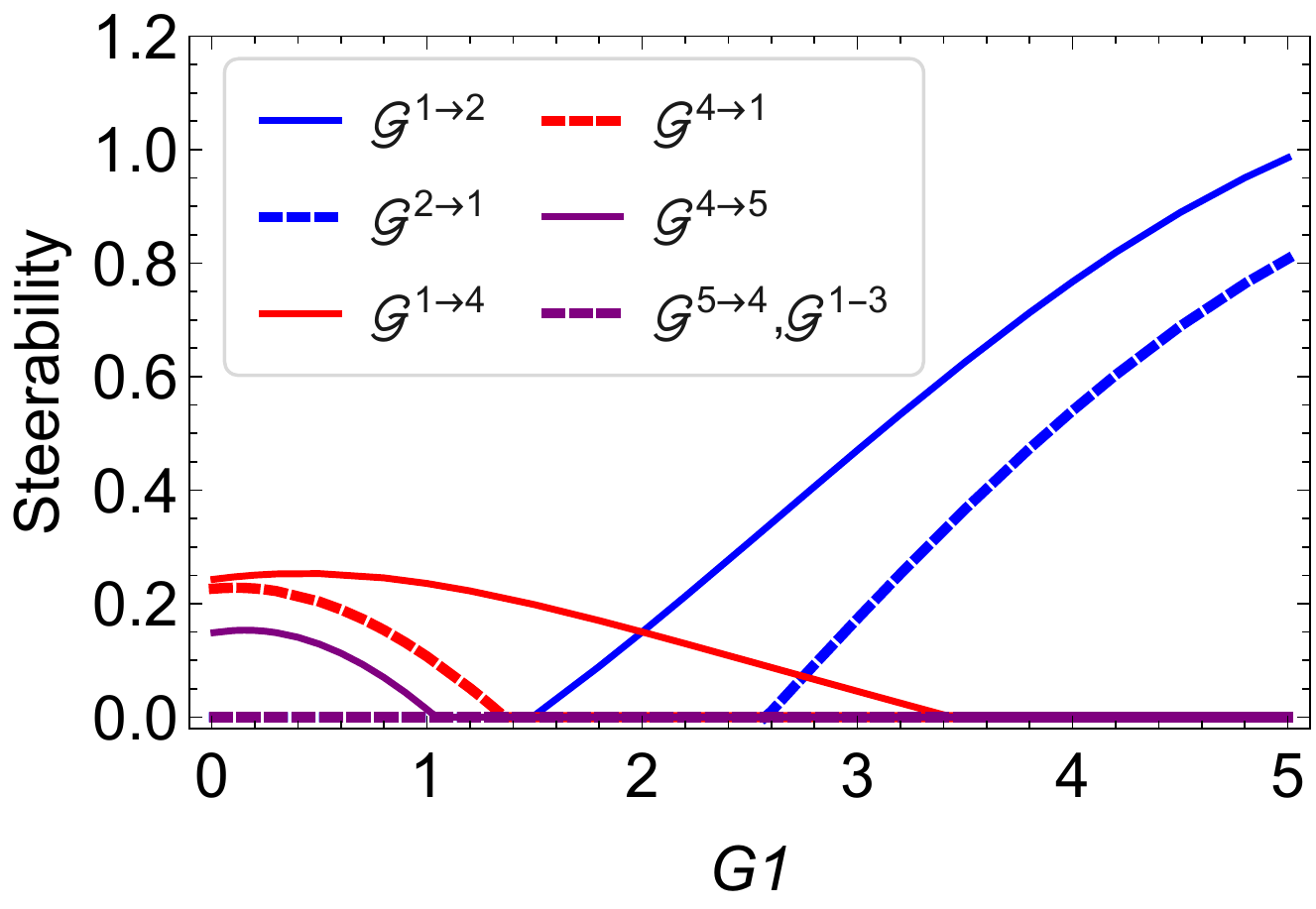}
\caption{The (1+1)-mode steerings. (a) The steerabilities versus the interaction time t when ${{G}_{2}}=1.2,{{G}_{3}}=2$. The (1+1)-mode steerings versus ${{G}_{1}}$ for $t=0.3$, (b) with ${{G}_{2}}=1.2,{{G}_{3}}=2$, and (c) with ${{G}_{2}}=2,{{G}_{3}}=1.2$.} 
\end{figure}

It is clear that steerings exist only between probe beams and conjugate beams. According to the symmetry in Fig.1(c), $\mathcal{G}^{1\to 2}=\mathcal{G}^{3\to 5}$, $\mathcal{G}^{1\to 4}=\mathcal{G}^{3\to 6}$, $\mathcal{G}^{5\to 4}=\mathcal{G}^{2\to 6}$, and vice verse. As is shown in Fig.2(b), one-way or two-way steering sensitively depends on the corresponding interaction strength. Steering between optical modes $\hat{a}_1$ and $\hat{a}_3$ (mainly depend on the double-pump interaction strength $G_3$) is always symmetric two-way steering, which can be well understood by their symmetry (both frequency and intensity), and will vanish when $G_1$ is bigger than 1.6, i.e., the critical interaction strength $G_1$ is between $G_2$ and $G_3$, which means $G_3$ should be stronger enough than the other two to obtain two-way steerings. For steering between $\hat{a}_1$ and $\hat{a}_2$, there only exist one-way steering from $\hat{a}_1$ to $\hat{a}_2$ when $2.6 < G_1 < 3.2$, and two-way asymmetric steering between $\hat{a}_1$ and $\hat{a}_2$ will happen when $G_1>3.2$. There exist a threshold because steering between $\hat{a}_1$ and $\hat{a}_2$ mainly depends on pump 1, which means $G_1$ should be strong enough. It is also noticed that steering from $\hat{a}_1$ to $\hat{a}_2$ is always stronger than that from $\hat{a}_2$ to $\hat{a}_1$, because all pump1, pump2 and double pump can affect mode $\hat{a}_1$ but only two of them (pump1 and double pump) can affect $\hat{a}_2$. For steering between $\hat{a}_4$ and $\hat{a}_5$, it mainly depends on double pump ($G_3$), however, $\hat{a}_4$ and $\hat{a}_5$ are also affected by pump2 and pump1, respectively. Therefore, when $G_1$ is small (<1.2), $\mathcal{G}^{4\to 5}>\mathcal{G}^{5\to 4}$, and when $G_1$ is bigger (>1.2), $\mathcal{G}^{4\to 5}<\mathcal{G}^{5\to 4}$, and when $G_1>2.1$ there only exist one-way steering from $\hat{a}_5$ to $\hat{a}_4$. It is interesting that symmetric two-way steering will happen at the intersection point $G_1=1.2$. In Fig.2(d), another parameters are chosen to show steerings between $\hat{a}_1$ and $\hat{a}_4$ clearly, which can be understood in the same way, and other steerings are also shown. 

\begin{figure}[h!]
\centering
\includegraphics[height=3.6cm,width=4.1cm]{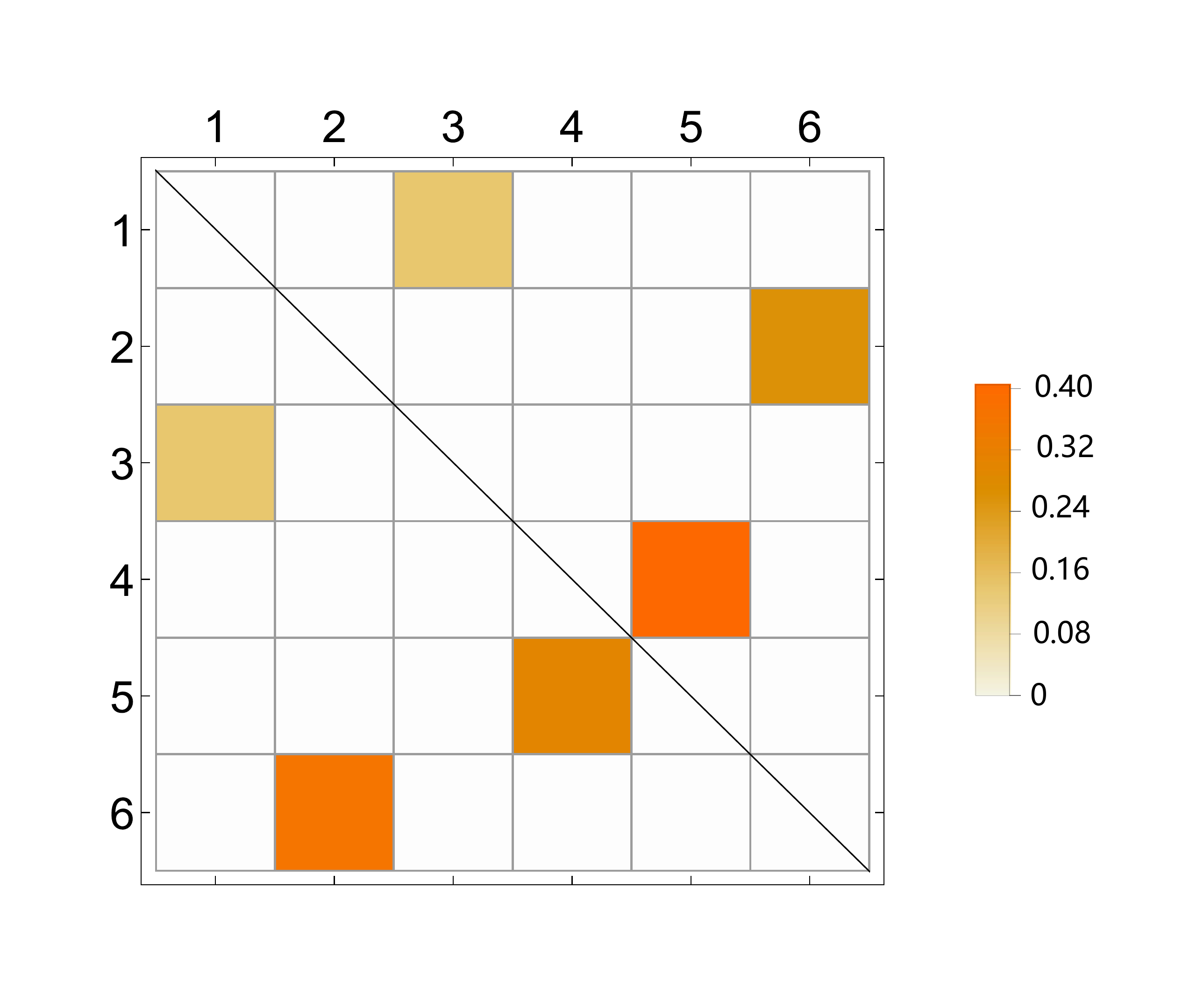} 
\hspace{0.03cm}
\includegraphics[height=3.6cm,width=4.1cm]{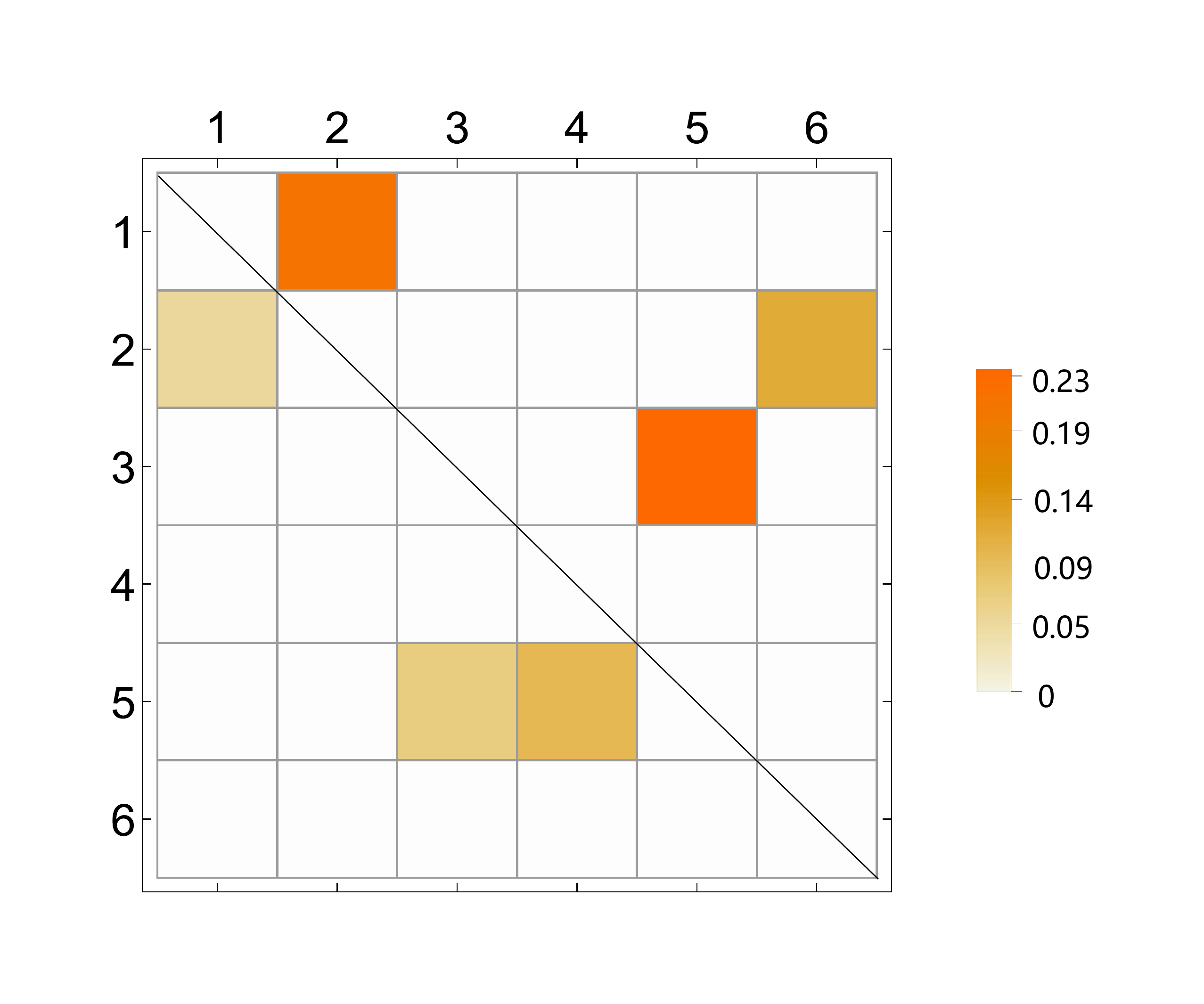}
\hspace{0.03cm}
\put(-193,99){\bf(a)}
\put(-70,99){\bf(b)}
\put(45,99){\bf(c)}
\includegraphics[height=3.6cm,width=4.1cm]{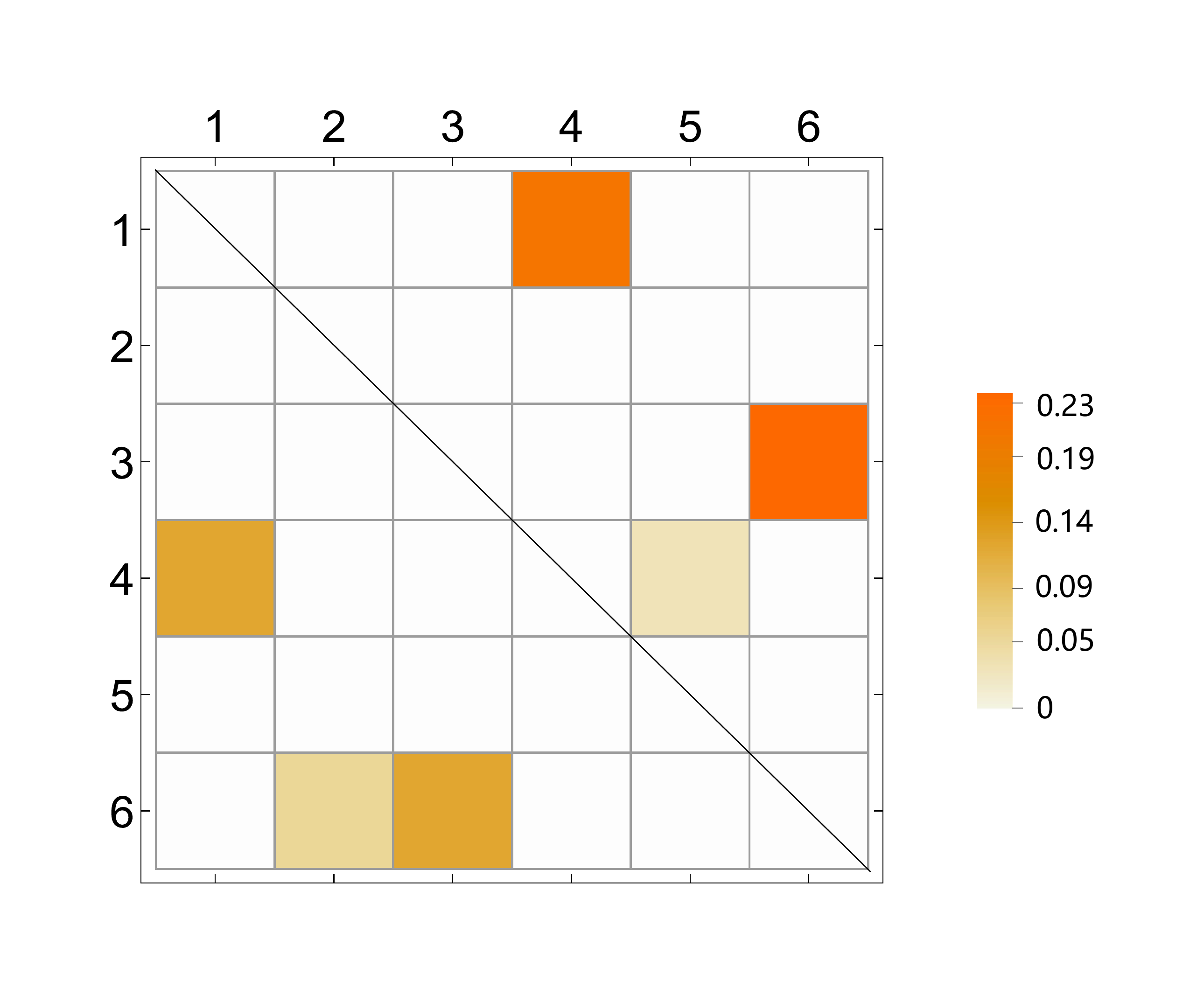}
\caption{
Matrix representation of (1+1)-mode  steerings. (a) ${{G}_{1}}=1, {{G}_{2}}=1.2, {{G}_{3}}=2$. (b) ${{G}_{1}}=3.5, {{G}_{2}}=1.2, {{G}_{3}}=2$.  (c) ${{G}_{1}}=1, {{G}_{2}}=2, {{G}_{3}}=1.2$.}. 
\end{figure}

Specially, matrix diagram can be used to represent all these one-way and two-way steerings, as is shown in Fig.3(a), Fig3(b) and Fig.3(c), with different parameters, respectively. Matrix element $A_{ij}$ represents the steering from i to j and different colors represent different steerabilities. Therefore, two-way steerings related to two elements that is symmetric to the diagonal line, while one-way steerings related to those elements that have no symmetric counterparts. For example, in Fig.3(a), three two-way steerings are demonstrated, however, only steering between $\hat{a}_1$ and  $\hat{a}_3$ is symmetric two-way (same color). This matrix representation can help us to demonstrate the steerings more intuitively. Furthermore, the satisfied type-I monogamy relations can also be found from these matrix representations, i.e., when $\hat{a}_1$ can be steered by $\hat{a}_3$, it cannot be steered by other modes simultaneously, ......, etc.

\subsection{The (1+2)/(2+1)-mode EPR steering} 
\begin{figure}[h!]
\centering
\includegraphics[height=2.8cm,width=3.7cm]{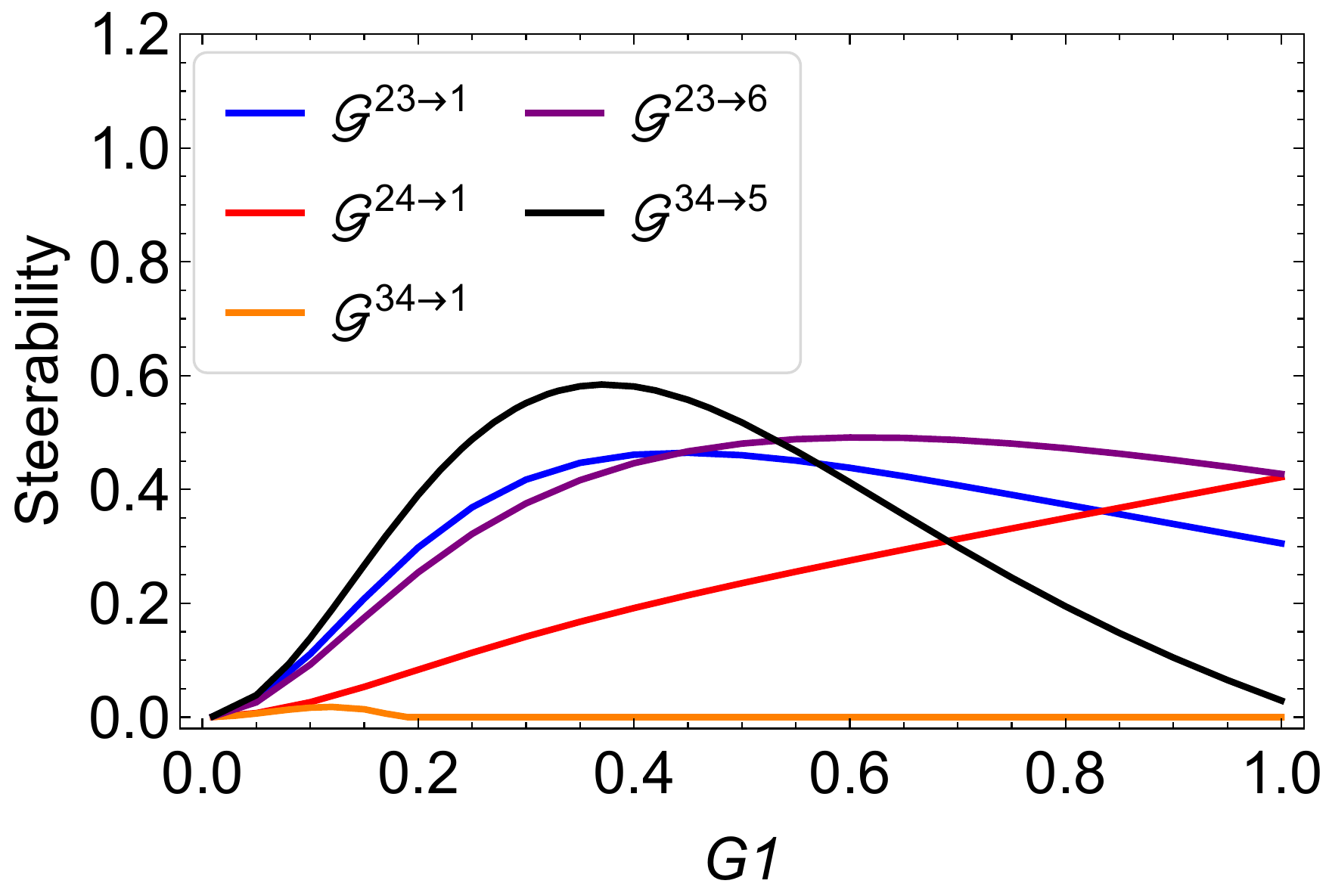}
\put(-53,85){\bf(a)}
\put(50,85){\bf(b)}
\put(153,85){\bf(c)}
\includegraphics[height=2.7cm,width=3.6cm]{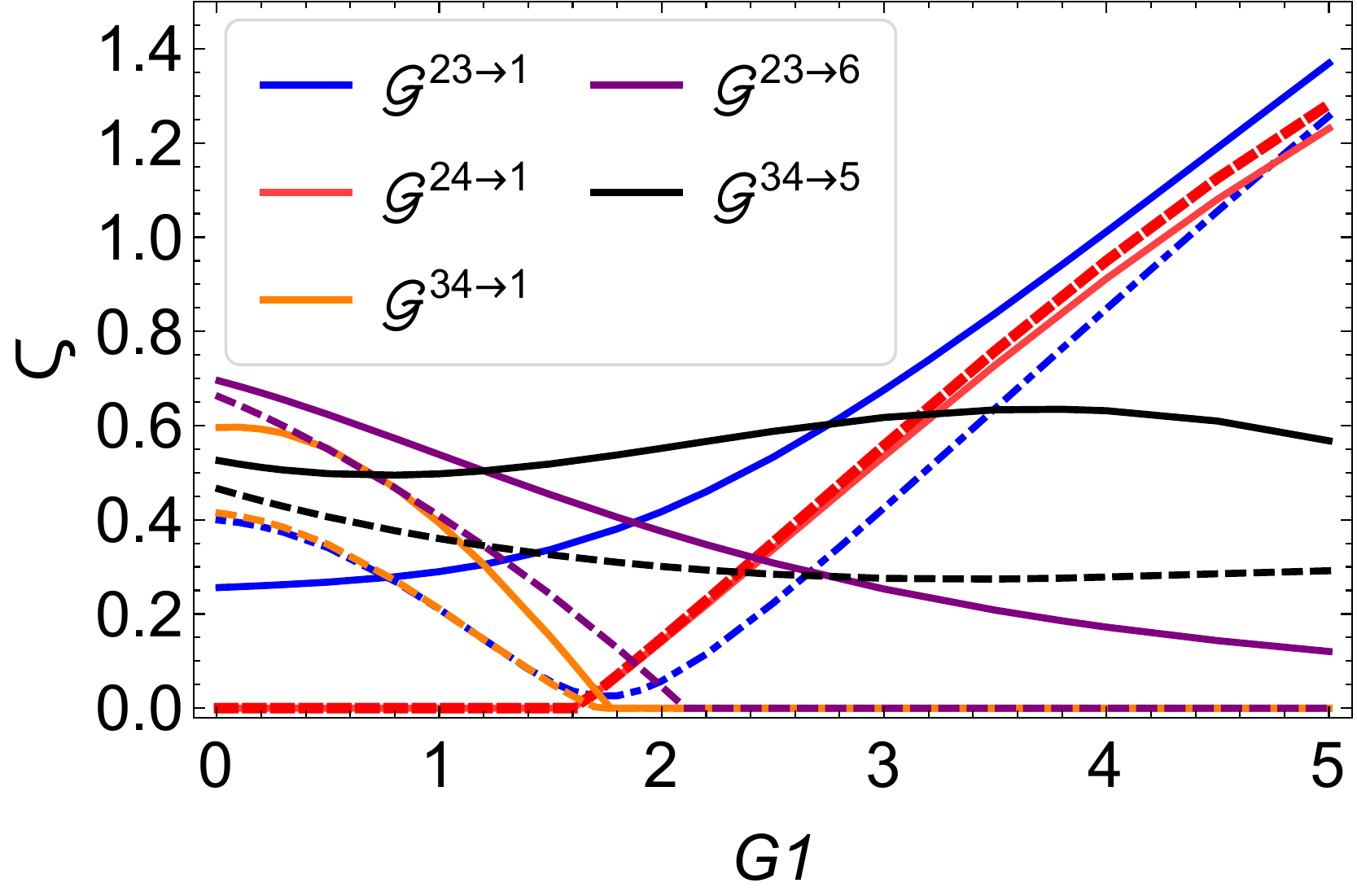}
\includegraphics[height=2.7cm,width=3.6cm]{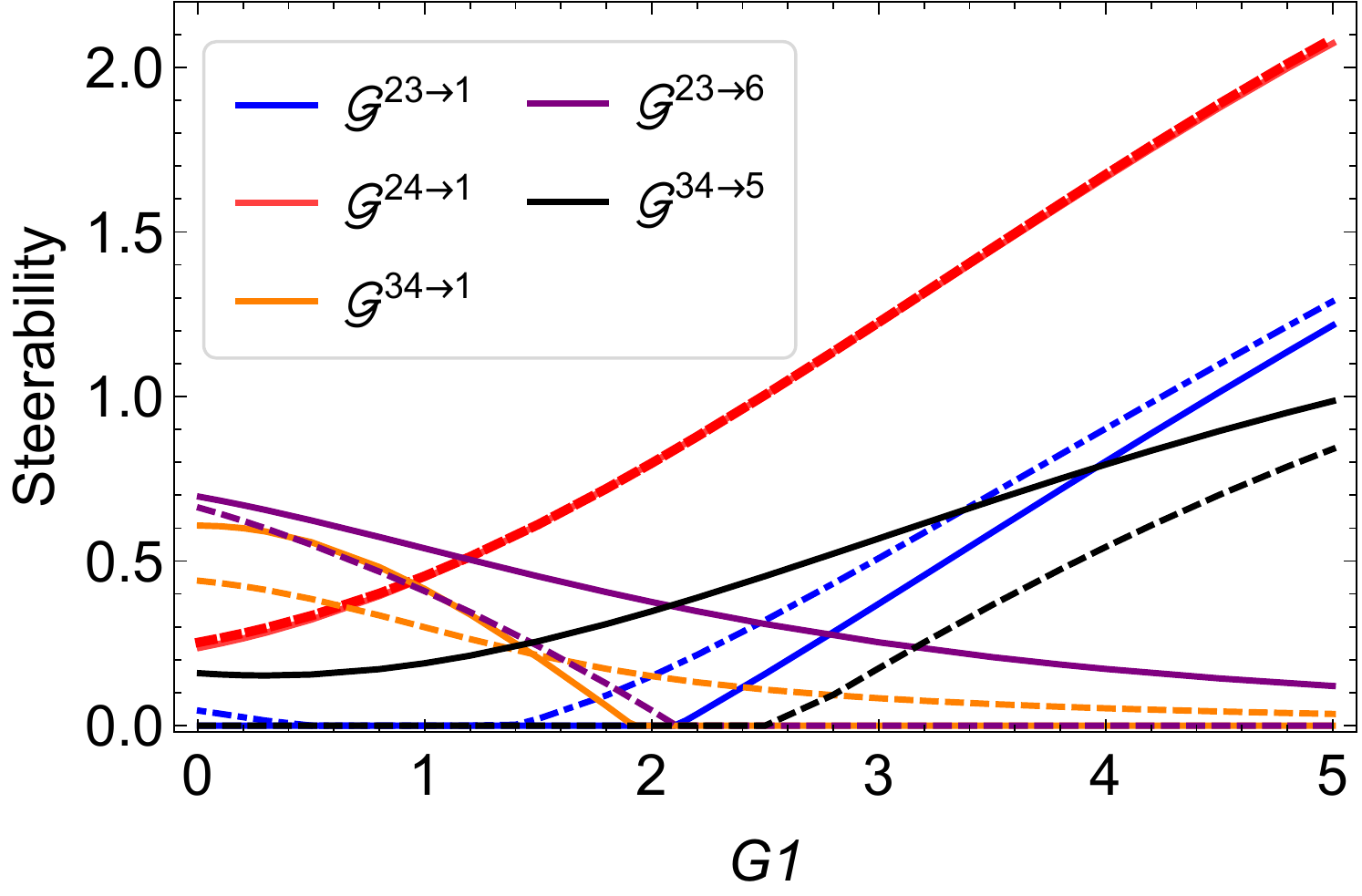}\\
\centering
\includegraphics[height=3.1cm,width=3.8cm]{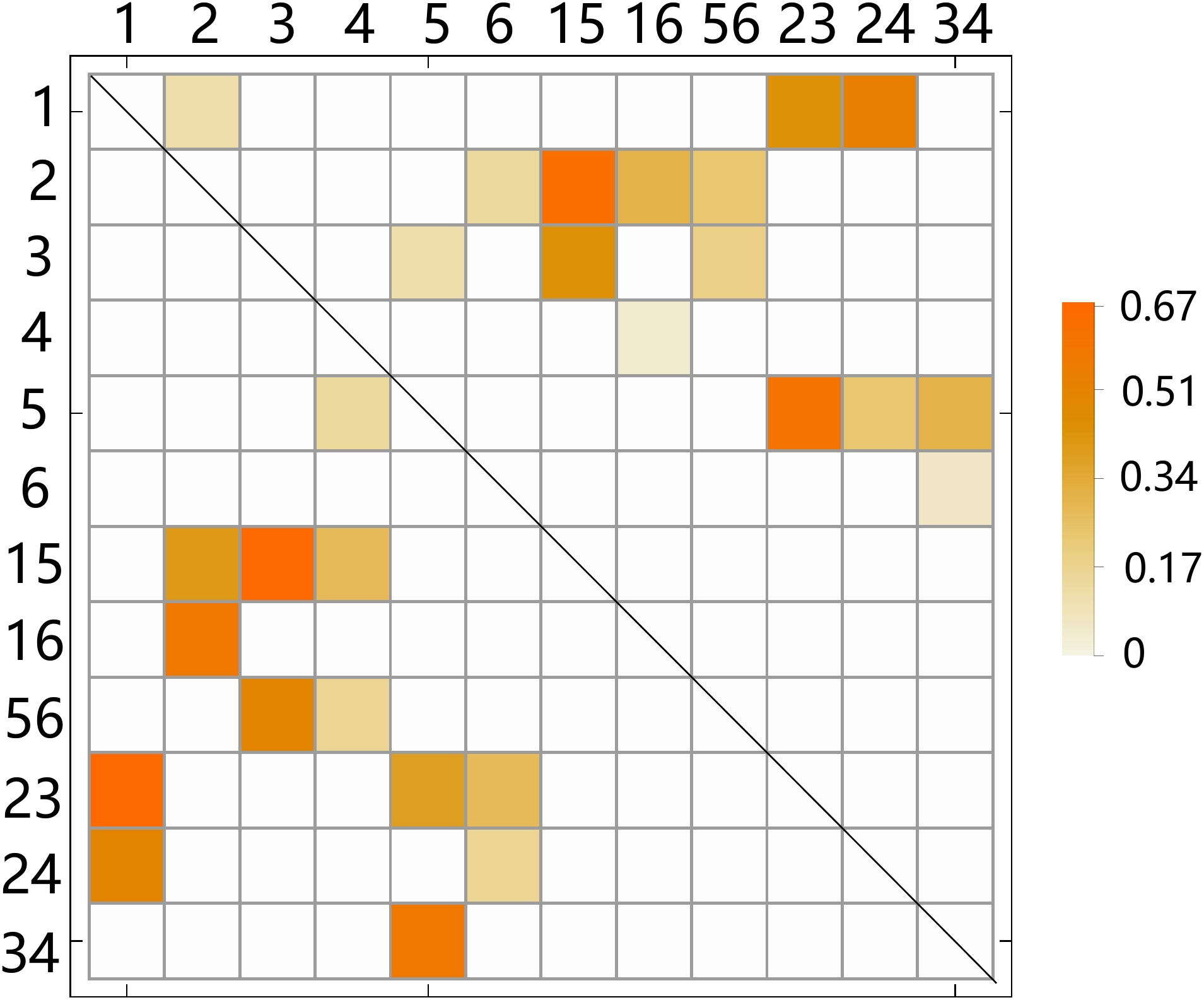}
\put(-67,90){\bf(d)}
\put(43,90){\bf(e)}
\includegraphics[height=3.1cm,width=3.8cm]{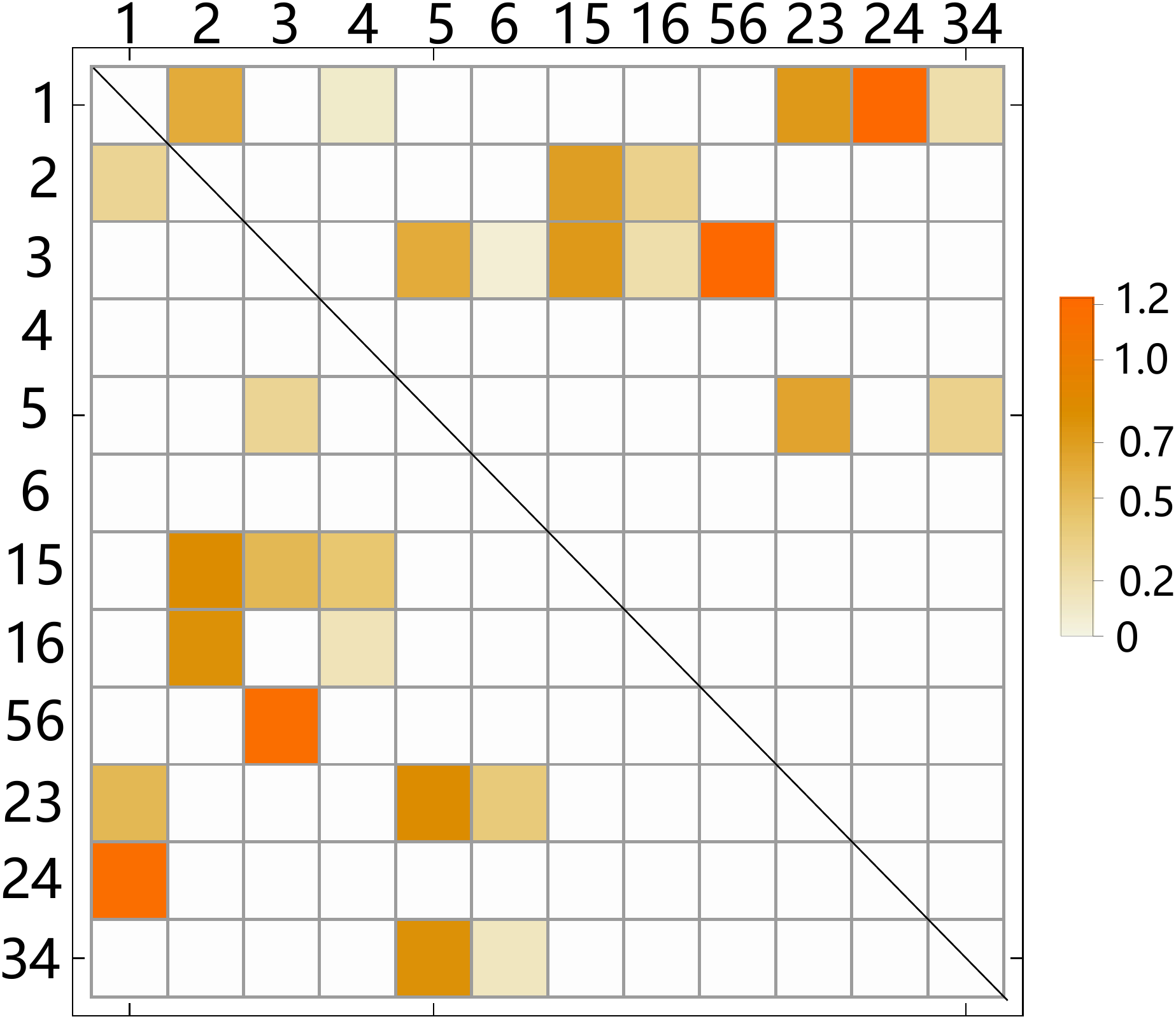}
\caption{ Results of the (1 + 2)/(2 + 1)-mode steerings. (a) Time evolution, with $G_1=2, G_3=2$, $G_2=1.2$. (b) The steerings versus $G_1$, $G_2=1.2, G_3=2$. (c) The steerings versus $G_1$, $G_2=2, G_3=1.2$. In subgraph (b) and (c), the solid line and dashed line of the same color correspond to a pair of counterparts, for example, the solid and dashed blue lines correspond to $\mathcal{G}^{23\to 1}$ and $\mathcal{G}^{1\to 23}$, respectively. (d) Matrix representation with $G_2=1.2, G_3=2$, $G_1=3$. (e) Matrix representation with $G_2=2, G_3=1.2$, $G_1=3$.}
\end{figure}
The (1+2)/(2+1)-mode steerings, which means the steered part or steering part contains two modes, are shown in Fig.4. $t=0.3$ is chosen for consistency, according to Fig.4(a). $\mathcal{G}^{34\to 5}=\mathcal{G}^{16\to 2}$, $\mathcal{G}^{23\to 6}=\mathcal{G}^{15\to 4}$, $\mathcal{G}^{24\to 1}=\mathcal{G}^{56\to 3}$, $\mathcal{G}^{23\to 1}=\mathcal{G}^{15\to 3}$, $\mathcal{G}^{34\to 1}=\mathcal{G}^{16\to 3}$, and vice verse, due to the symmetry in Fig.1(c). As is shown in Fig.4(b), steering between modes $\hat{a}_2\hat{a}_3$ and mode $\hat{a}_6$ will change from two-way to one-way when $G_1>2.1$, and steerings will decrease when $G_1$ increase because they are mainly depend on the interaction strength $G_2$ and $G_3$. For steering between modes $\hat{a}_2\hat{a}_3$ and mode $\hat{a}_1$ is always two-way, when $G_1$ is small (<0.8), $\mathcal{G}^{1\to 23}>\mathcal{G}^{23\to 1}$, and when $G_1$ is bigger (>0.8), $\mathcal{G}^{1\to 23}<\mathcal{G}^{23\to 1}$. It is interesting that symmetric two-way steering will happen at the intersection point $G_1=0.8$, and both steerings will increase when $G_1$ is further increasing because they are mainly depend on $G_1$ and $G_3$. Steering between modes $\hat{a}_3\hat{a}_4$ and mode $\hat{a}_5$ is always two-way due to strong $G_3$. Steering between modes $\hat{a}_3\hat{a}_4$ and mode $\hat{a}_1$ is two-way and will vanish when $G_1>1.8$ due to smaller relative strength of $G_2$. Steering between modes $\hat{a}_2\hat{a}_4$ and mode $\hat{a}_1$ has a smaller threshold compare with that between $\hat{a}_2$ and $\hat{a}_1$(shown in Fig.2(b)) due to the contribution of pump2. Results for bigger relative strength of $G_2$ are shown in Fig.4(c). In Fig.4(c), Steering between modes $\hat{a}_2\hat{a}_4$ and mode $\hat{a}_1$ is always two-way due to stronger $G_2$, and Steering between modes $\hat{a}_3\hat{a}_4$ and mode $\hat{a}_5$ will change from one-way to two-way  $G_1>2.5$, and Steering between modes $\hat{a}_3\hat{a}_4$ and mode $\hat{a}_1$ will change from two-way to one-way when $G_1>1.9$ with a intersection of symmetric two-way steering at $G_1=1.5$. Fig.4(d) and Fig.4(e) are the matrix representations under different parameters, which contain all possible (1+2)/(2+1) modes. From the matrix representations, it is found that the type-II monogamy relations also can be satisfied, for example, when $\hat{a}_5$ can be steered by $\hat{a}_2\hat{a}_3$, it cannot be steered by $\hat{a}_4$ simultaneously; when $\hat{a}_1$ can be steered by $\hat{a}_2\hat{a}_4$, it cannot be steered by $\hat{a}_3$ simultaneously; .......  

\subsection{The (1+3)/(3+1)-mode EPR steering}
\begin{figure}[htbp]
\centering
\quad
\put(55,90){\bf(a)}
\put(170,90){\bf(b)}
\put(280,90){\bf(c)}
\centering
\includegraphics[height=3cm,width=4.cm]{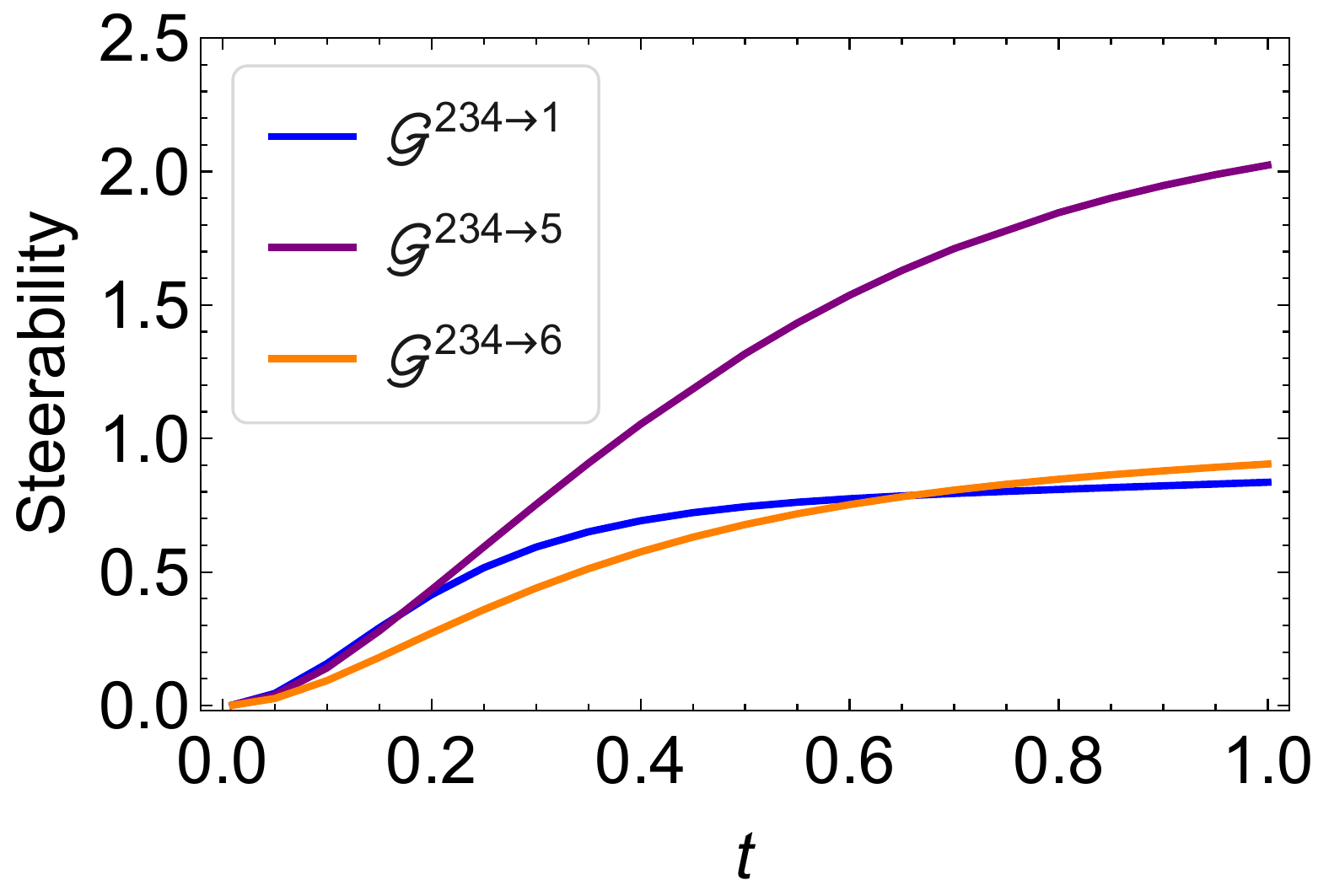}
\includegraphics[height=3cm,width=3.85cm]{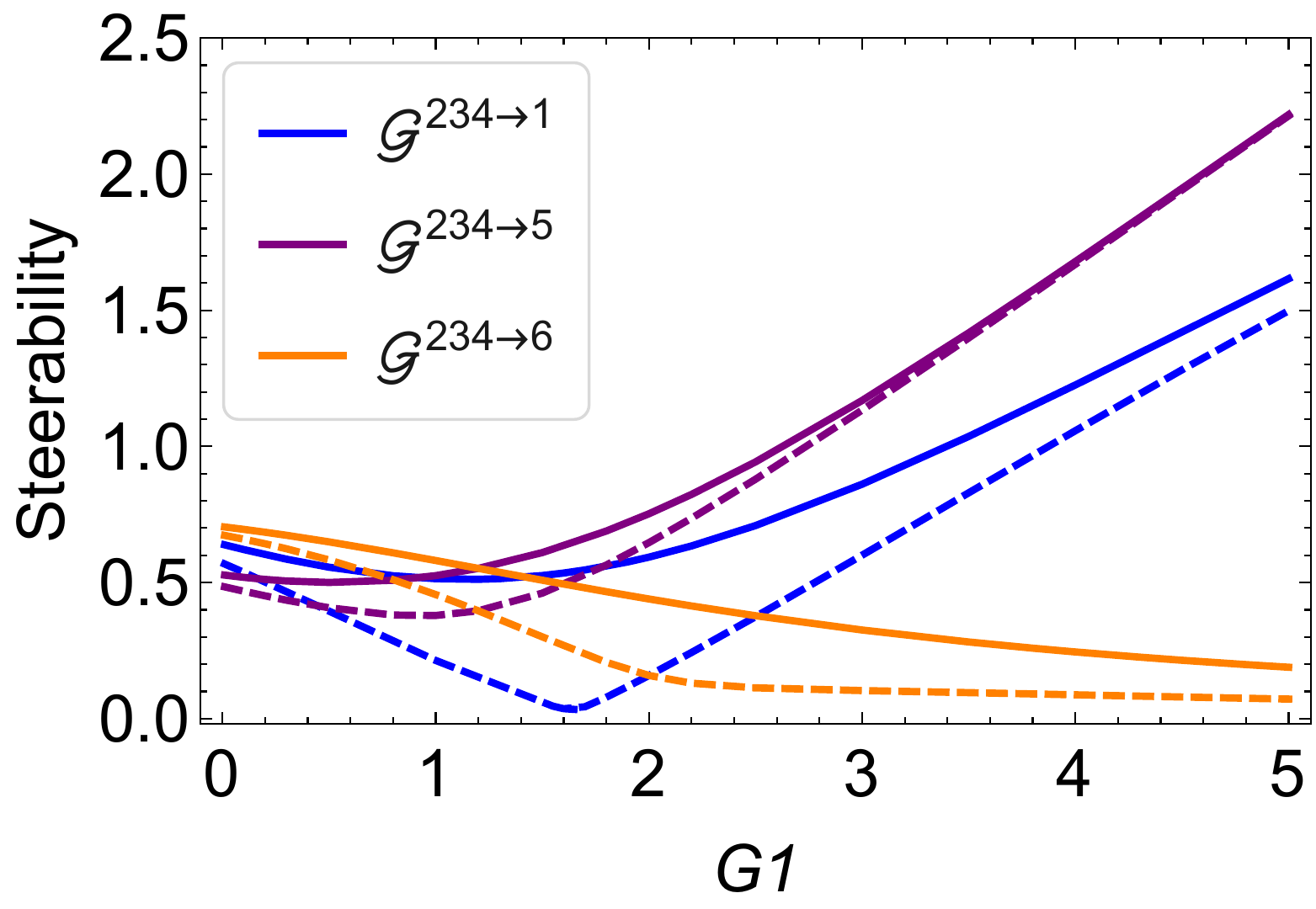}
\includegraphics[height=3cm,width=3.85cm]{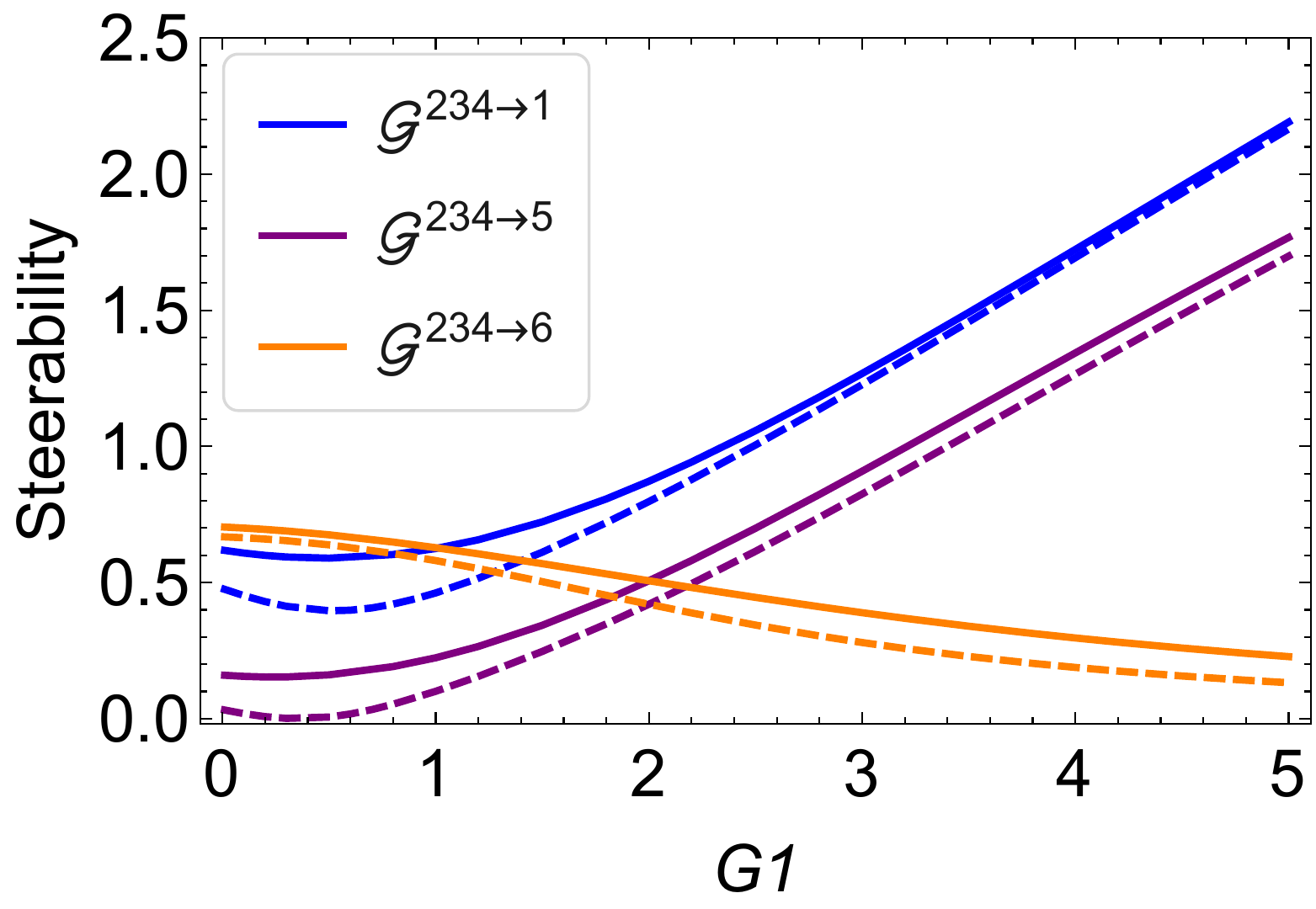}\\
\centering
\includegraphics[height=3.1cm,width=4.2cm]{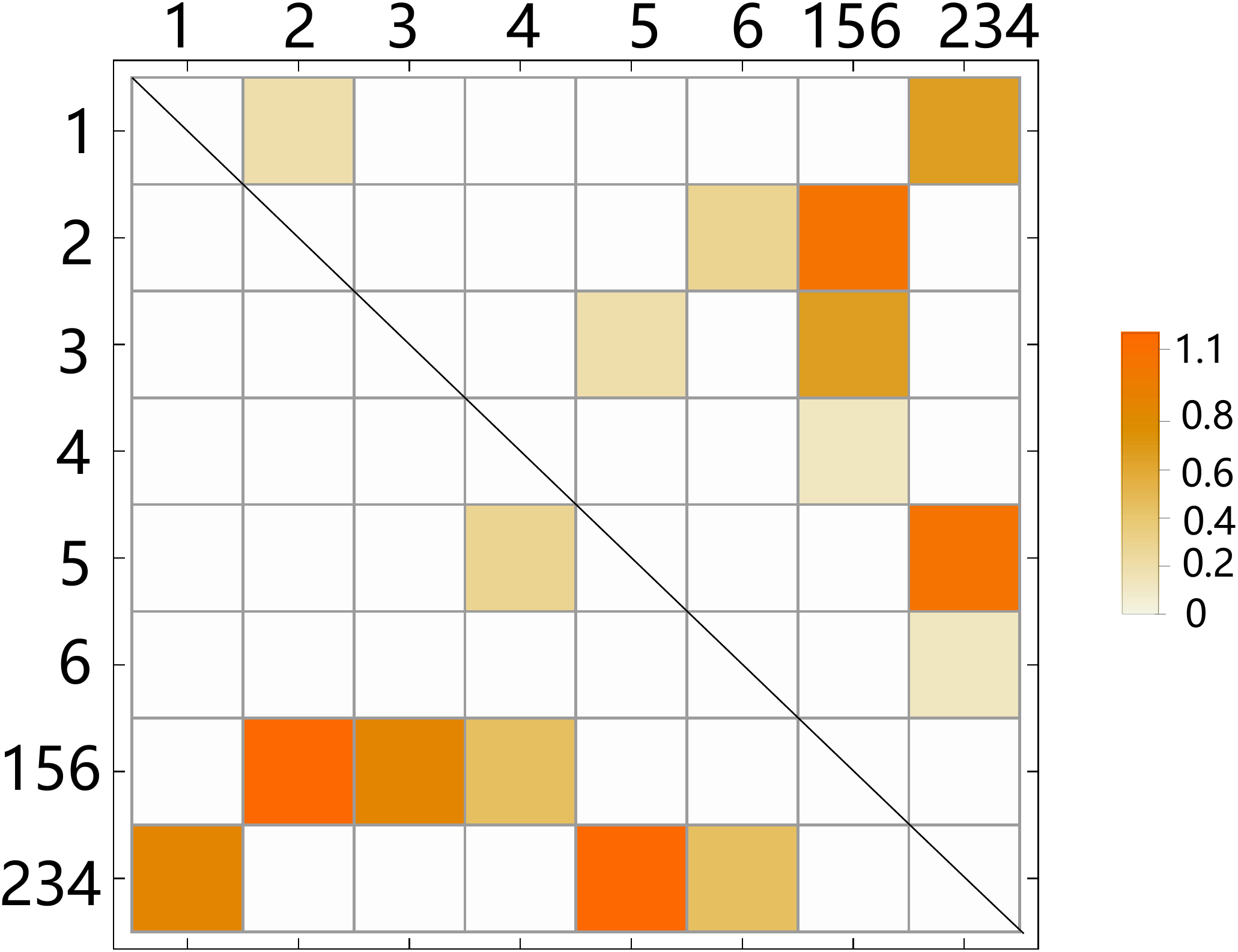}
\put(-70,93){\bf(d)}
\put(50,93){\bf(e)}
\includegraphics[height=3.1cm,width=4.2cm]{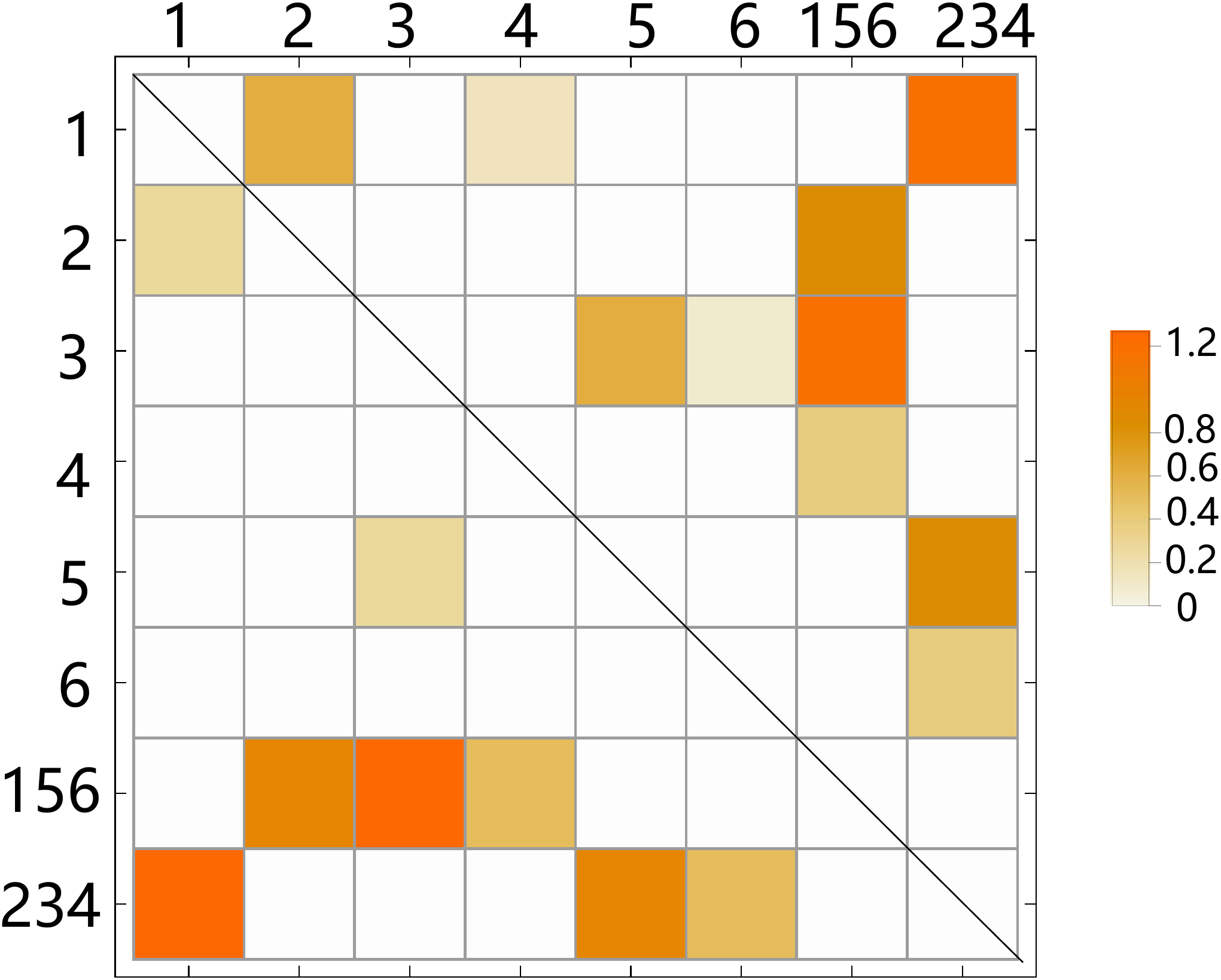}
\caption{The (1+3)/(3+1)-mode steerings. (a) The time evolution diagram with $G_1=2, G_3=2$, $G_2=1.2$. (b) The steerabilities versus $G_1$, where $G_2=1.2, G_3=2$. (c) The steerabilities versus $G_1$, where $G_2=2, G_3=1.2$. In subgraph (b) and (c), the solid line and dashed line of the same color correspond to a pair of counterparts, for example, the solid and dashed blue lines correspond to $\mathcal{G}^{234\to 1}$ and $\mathcal{G}^{1\to 234}$, respectively. (d) Matrix representation diagram when  $G_1=3, G_2=1.2, G_3=2$; (e) $G_1=3, G_2=2, G_3=1.2$.}      
\end{figure}
The (1+3)/(3+1)-mode steerings are shown in Fig.5. t=0.3 is still chosen for consistency based on the time evolution diagram Fig.5(a), and it can be found that the steerabilities are robust and stronger when the steering part or the steered part contain three modes. Still we have $\mathcal{G}^{234\to 1}=\mathcal{G}^{156\to 3}$, $\mathcal{G}^{234\to 5}=\mathcal{G}^{156\to 2}$, $\mathcal{G}^{234\to 6}=\mathcal{G}^{156\to 4}$, and vice verse, due to the symmetry in Fig.1(c). Specially, all (1+3)/(3+1)-mode steerings are asymmetric two-way, as is shown in Fig.5(b) and Fig.5(c). It means the more the steering mode, the less dependence on the relative strength. However, bigger relative strength of $G_2$ leads to stronger steering between modes $\hat{a}_2 \hat{a}_3\hat{a}_4$ and mode $\hat{a}_1$, compared with that between $\hat{a}_2 \hat{a}_3\hat{a}_4$ and $\hat{a}_5$. With the increase of $G_1$, $\mathcal{G}^{1\to 234}$ decrease, while $\mathcal{G}^{234\to 1}$ and $\mathcal{G}^{234\to 5}$ finally increase. Generally, steering from one-mode to multi-modes is smaller than that of multi-mode to one-mode, for example, $\mathcal{G}^{234\to 1}>\mathcal{G}^{1\to 234}$ (The solid line is higher than the dashed line of the same color). It is clearly shown from the matrix representation in Fig.5(d) and Fig.5(e), all (1+3)/(3+1)-mode steerings are asymmetric two-way steerings. Moreover, from the matrix representations, it is found that the type-III monogamy relations can be satisfied, for example, the steeribility $\mathcal{G}^{156\to 2}$ is bigger than the sum steeribilities $\mathcal{G}^{1\to 2}$, $\mathcal{G}^{5\to 2}$, and $\mathcal{G}^{6\to 2}$; the steeribility of $\mathcal{G}^{2\to 156}$ is bigger than the sum of steeribilities $\mathcal{G}^{2\to 1}$, $\mathcal{G}^{2\to 5}$, and $\mathcal{G}^{2\to 6}$; .......  

\section{Collective multipertite EPR steering}

For a quantum system consisting of $n$ parties, if a given party $i$ can be steered by all the remaining $n-1$ parties, but not by any $n-2$ parties, it is called collective multipartite steering. It means that the group of $n-1$ parties must collaborate after performing local measurements on each individual system in order to extract the information of party $i$, which has potential application to achieve ultra-secure n-party quantum secret sharing\cite{He2014Co,Yin2021}.  And this kind of special quantum states may provide a solution to the challenges of secure quantum communication network.
\begin{table}[h!]
\centering
\begin{center}
\caption{All possible collective steering}
	\begin{tabular}{ | m{2cm}<{\centering}|m{2cm}<{\centering} | m{2cm}|}
		\hline
\centering
 \begin{minipage}[b]{0.2\columnwidth}
		\raisebox{-.5\height}{\includegraphics[width=0.7\linewidth]{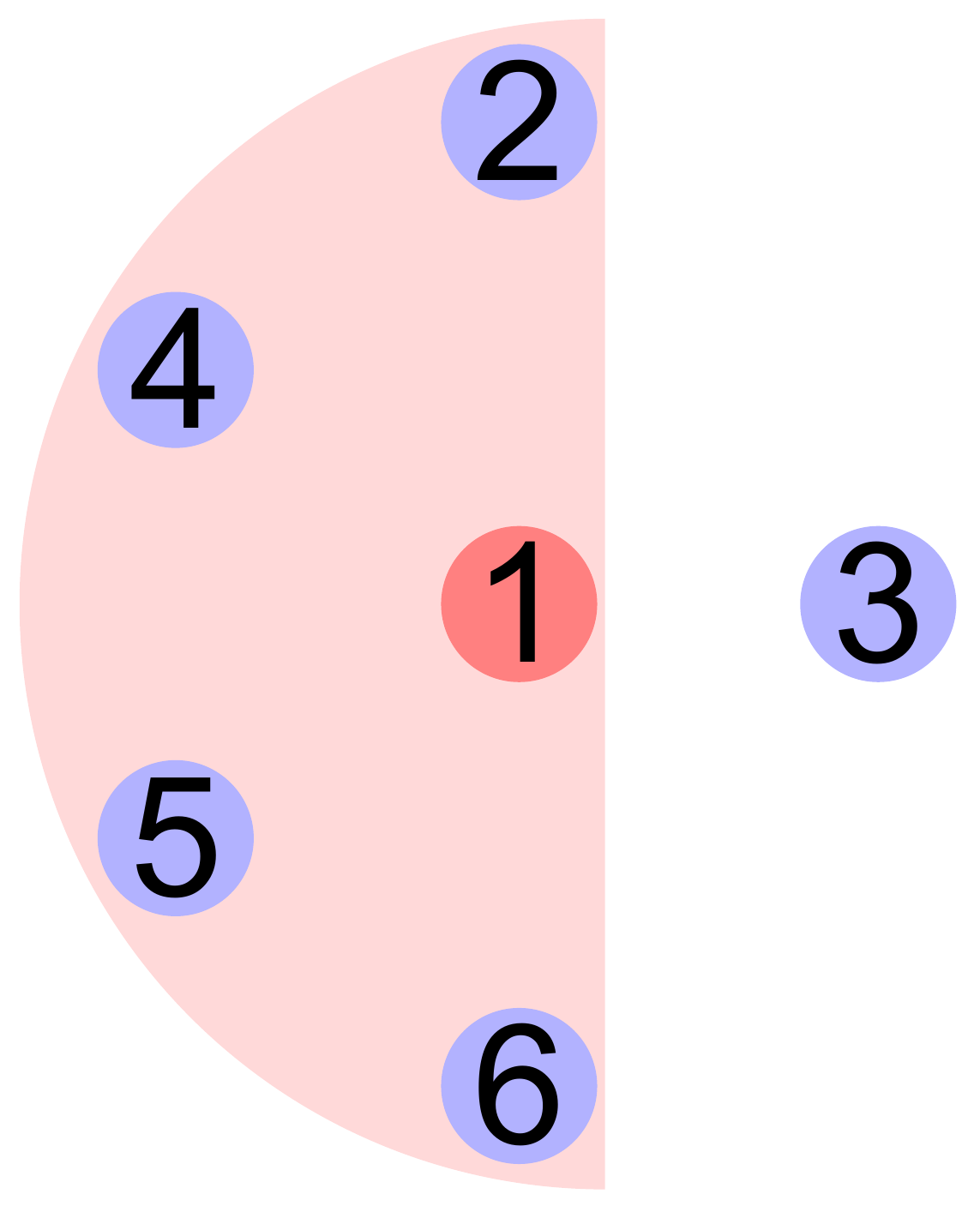}}
	\end{minipage}  & \begin{minipage}[b]{0.2\columnwidth}
		\raisebox{-.5\height}{\includegraphics[width=0.7\linewidth]{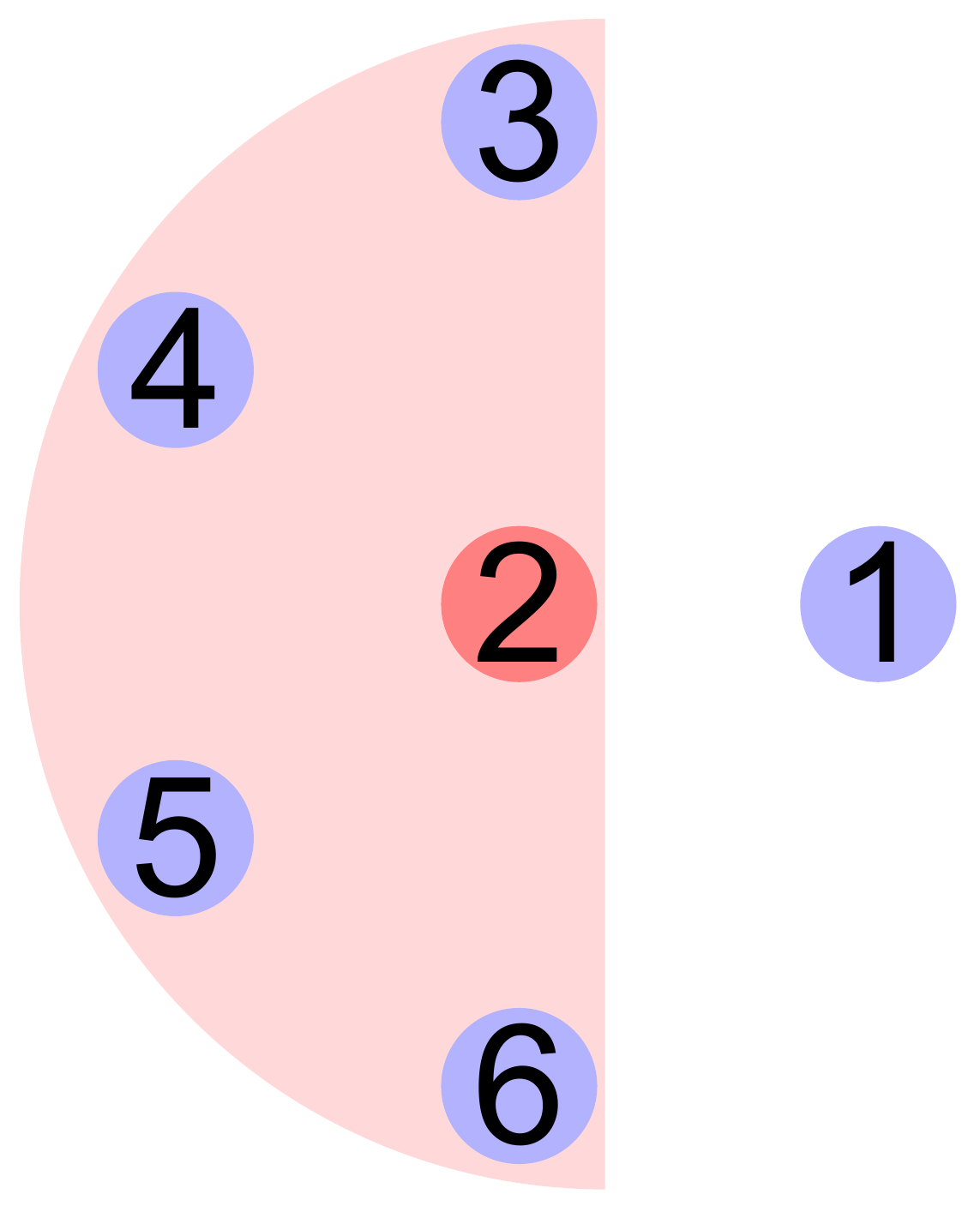}}
	\end{minipage} & \begin{minipage}[b]{0.2\columnwidth}
		\raisebox{-.5\height}{\includegraphics[width=0.7\linewidth]{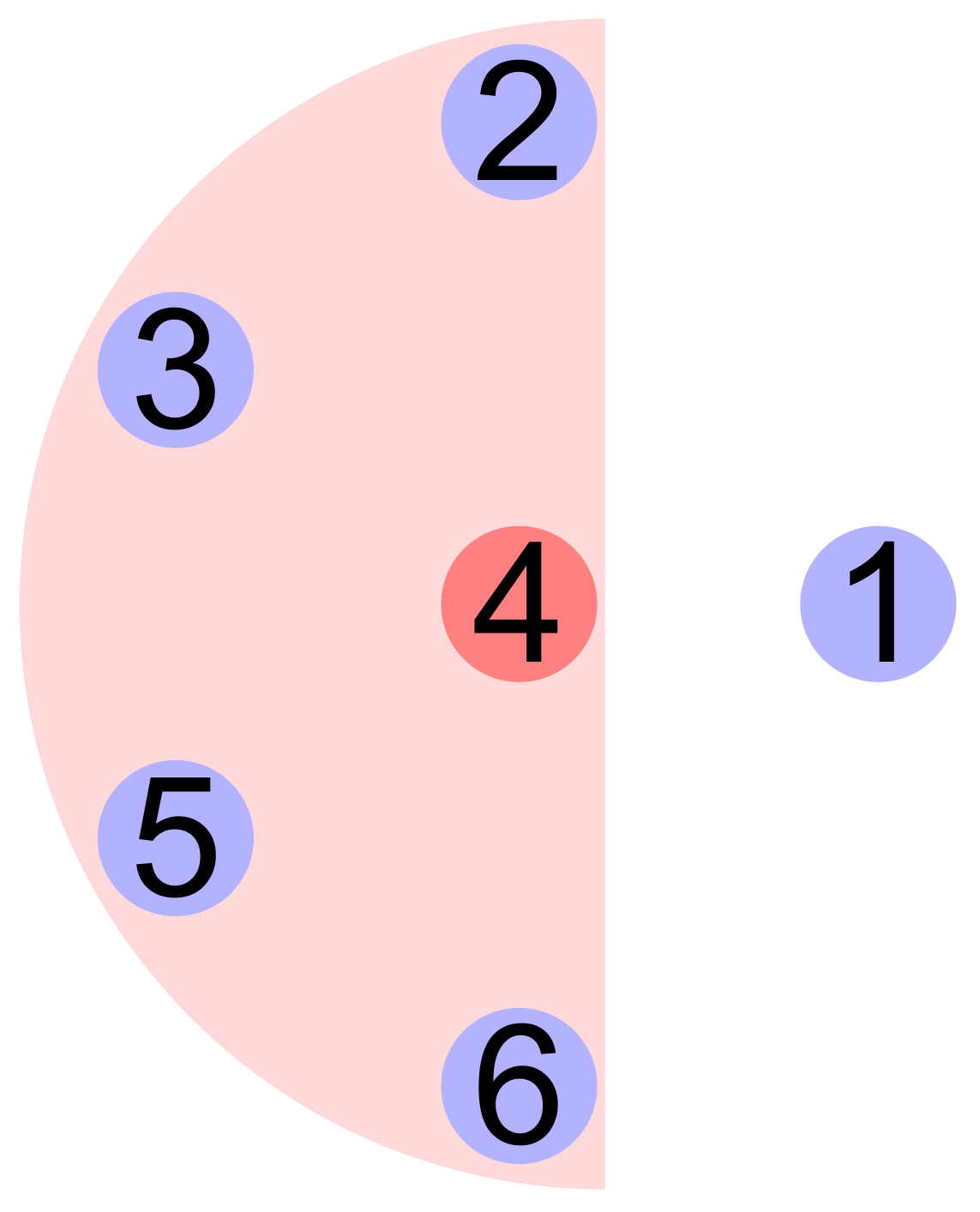}}
	\end{minipage} \\ \hline
 \centering
  \begin{minipage}[b]{0.2\columnwidth}
		\raisebox{-.5\height}{\includegraphics[width=0.7\linewidth]{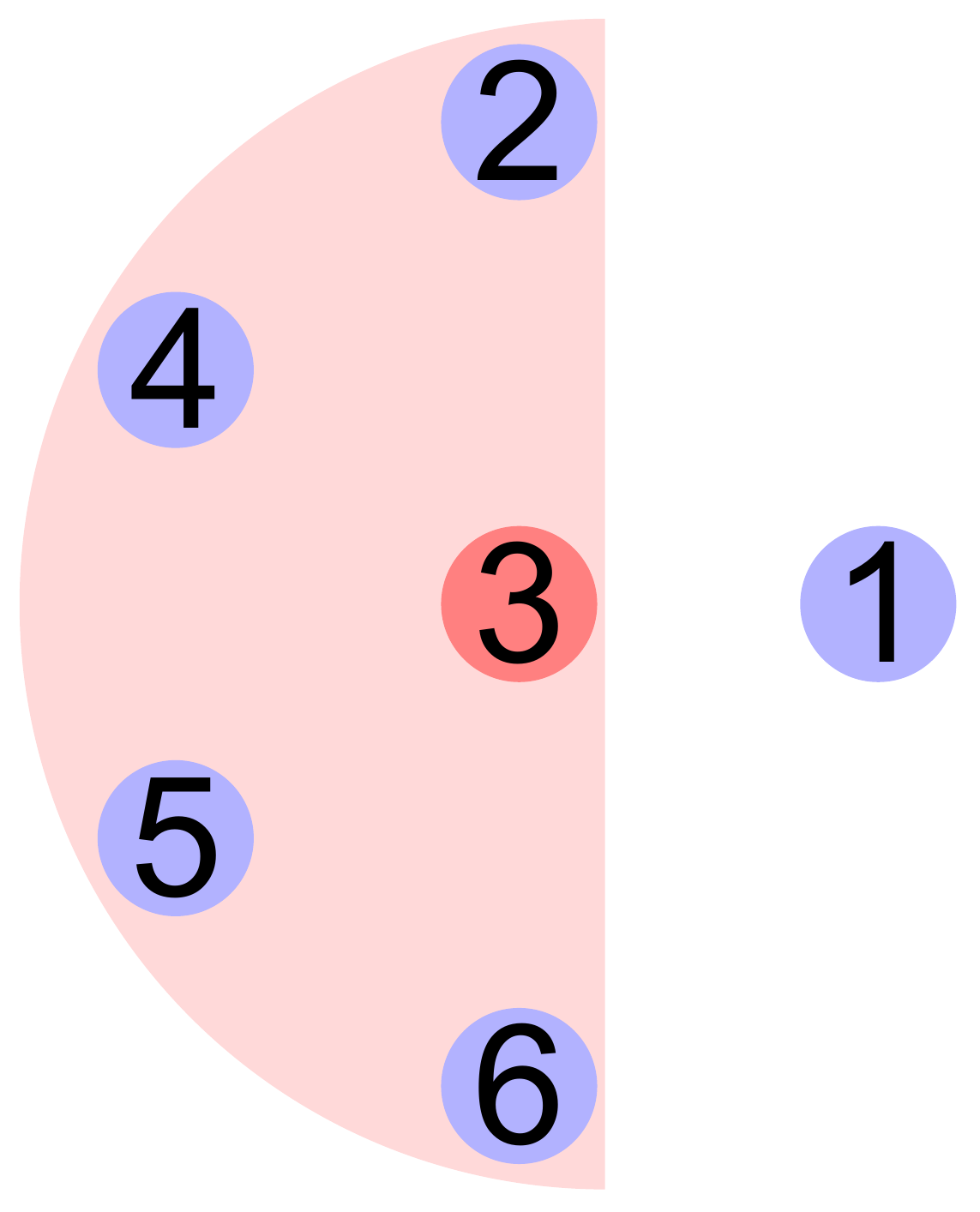}}
	\end{minipage} & \begin{minipage}[b]{0.2\columnwidth}
		\raisebox{-.5\height}{\includegraphics[width=0.7\linewidth]{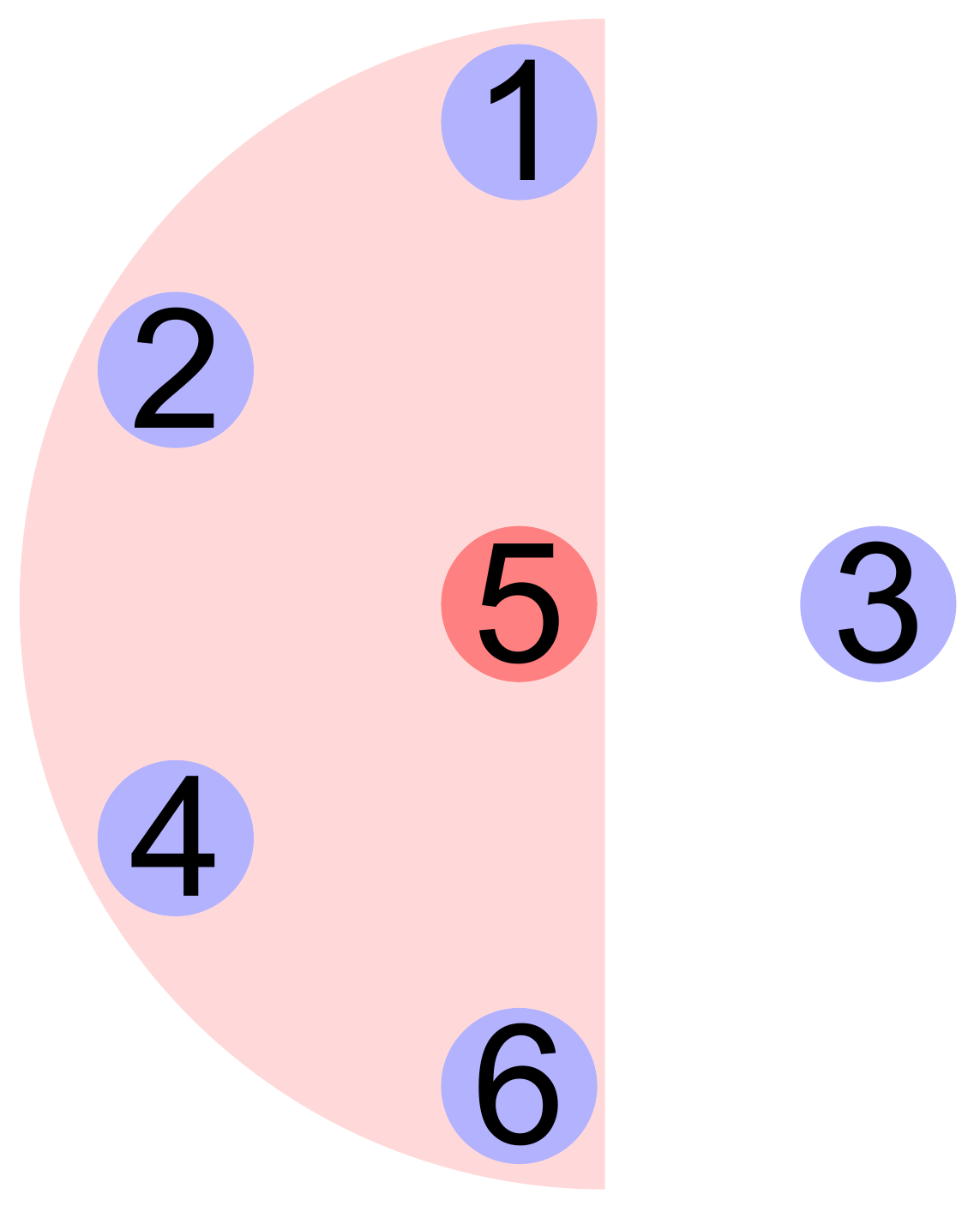}}
	\end{minipage} & \begin{minipage}[b]{0.2\columnwidth}
		\raisebox{-.5\height}{\includegraphics[width=0.7\linewidth]{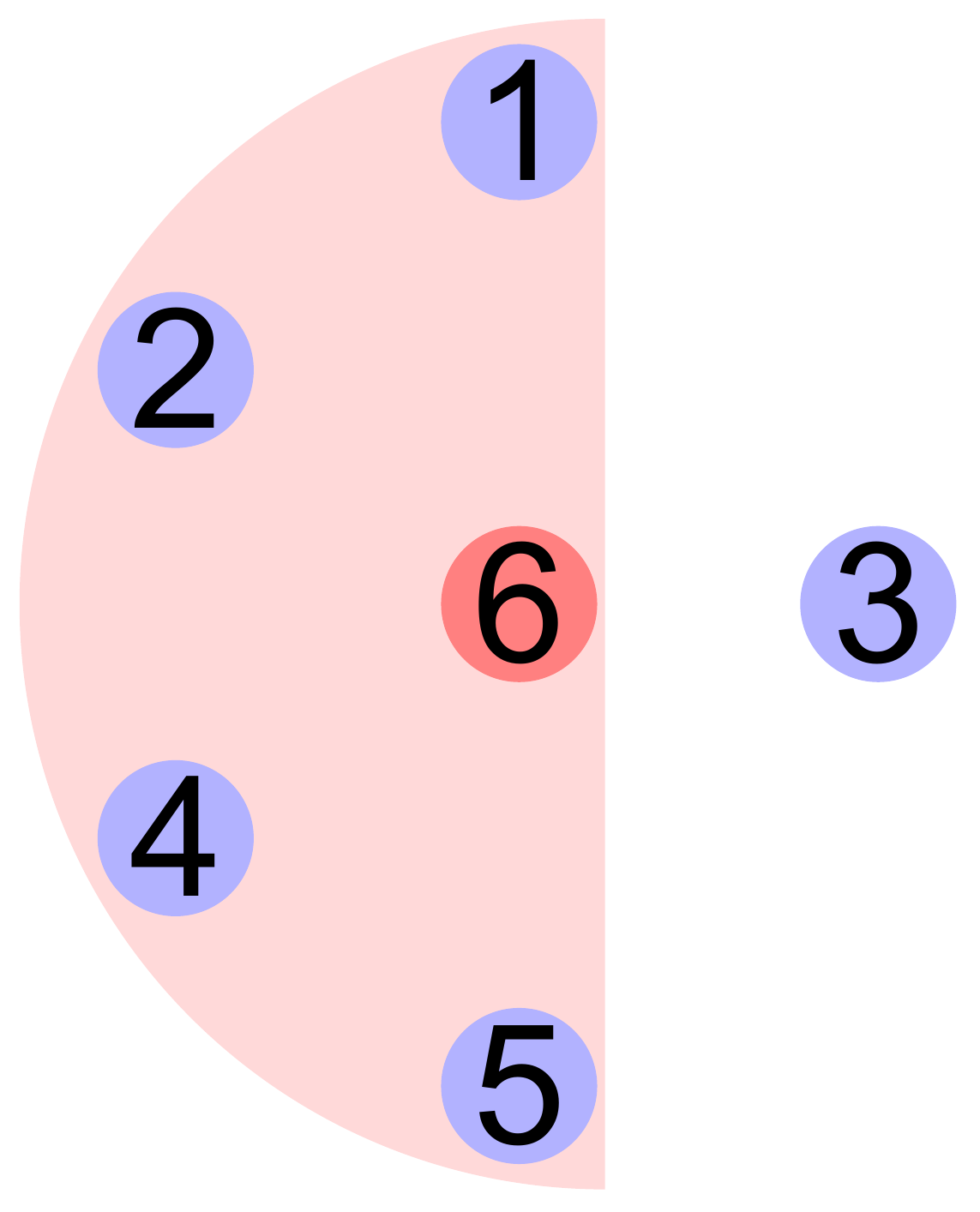}}
	\end{minipage} \\ \hline
\end{tabular}
\end{center}
\end{table}

Considering a n-partite case that has one steered mode ($B$) and $n-1$ steering modes ($A_1, A_2,...,A_{n-1}$), the criterion of the collective multipartite steering can be simplified as\cite{Yin2021}

\begin{equation}
    \begin{aligned}\label{eq7}
   \mathcal{G}^{A^\prime \to B} &=0, \quad  {n_{A^\prime}} = n-2 \\
   \mathcal{G}^{A_1A_2...A_{n-1}\to B} &>0, \quad  n_A = n-1\\   
    \end{aligned}
\end{equation}
Here $A^\prime$ includes all $C_{n-1}^{1}$possibilities of one less mode than $A$ with $n-1$ mode. It is noticed that only steering party with $n-1$ modes can steer mode $B$ but $n-2$ modes cannot. For six output modes in our model, there exist collective pentapartite steering at most and all its possibilities are shown in Stable 1, among which there are three conditions of collective steerings according to three columns, i.e., the two structures in each column have the same steeribility, due to the symmetry. Collective pentapartite EPR steering including modes $(\hat{a}_1 \hat{a}_2 \hat{a}_4 \hat{a}_5 \hat{a}_6)$ and $(\hat{a}_2 \hat{a}_3 \hat{a}_4 \hat{a}_5 \hat{a}_6)$ based on the above three conditions are shown in Fig.6. It is clear in Fig.6(a), for pentapartite modes $(\hat{a}_1 \hat{a}_2 \hat{a}_4 \hat{a}_5 \hat{a}_6)$, if $G_3>4.3$ and $G_1=1,G_2=3.2$ is chosen, collective steering that only $\mathcal{G}^{2456\to 1}$ can happen and steerings of less steering modes vanish, i.e., $\mathcal{G} ^{2456\to 1}>0, \mathcal{G} ^{256\to 1}=\mathcal{G}^{456\to 1}=\mathcal{G}^{246\to 1}=\mathcal{G} ^{245\to 1}=0$. 
For pentapartite modes $(\hat{a}_2 \hat{a}_3 \hat{a}_4 \hat{a}_5 \hat{a}_6)$, if the steered mode is $\hat{a}_2$ and $G_1=4,G_2=2$ is chosen, the collective steering ($\mathcal{G} ^{3456\to 2}>0, \mathcal{G} ^{345\to 2}=\mathcal{G} ^{356\to 2}=\mathcal{G} ^{346\to 2}=\mathcal{G} ^{456\to 2}=0$) will happen when $2.4<G_3<3.0$, as is shown in Fig.6(b). While if the steered mode is $a_4$ and $G_1=1.5,G_2=4$ is chosen, the collective steering ($\mathcal{G} ^{2356\to 4}>0, \mathcal{G} ^{235\to 2}=\mathcal{G} ^{236\to 2}=\mathcal{G} ^{356\to 2}=\mathcal{G} ^{256\to 2}=0$) will happen when $2.45<G_3<3.42$, as is shown in Fig.6(c).
\begin{figure}[htbp]
\centering
\quad
\put(55,95){\bf($a$)}
\put(175,95){\bf($b$)}
\put(290,95){\bf($c$)}
\centering
     \includegraphics[height=3.3cm,width=4.1cm]{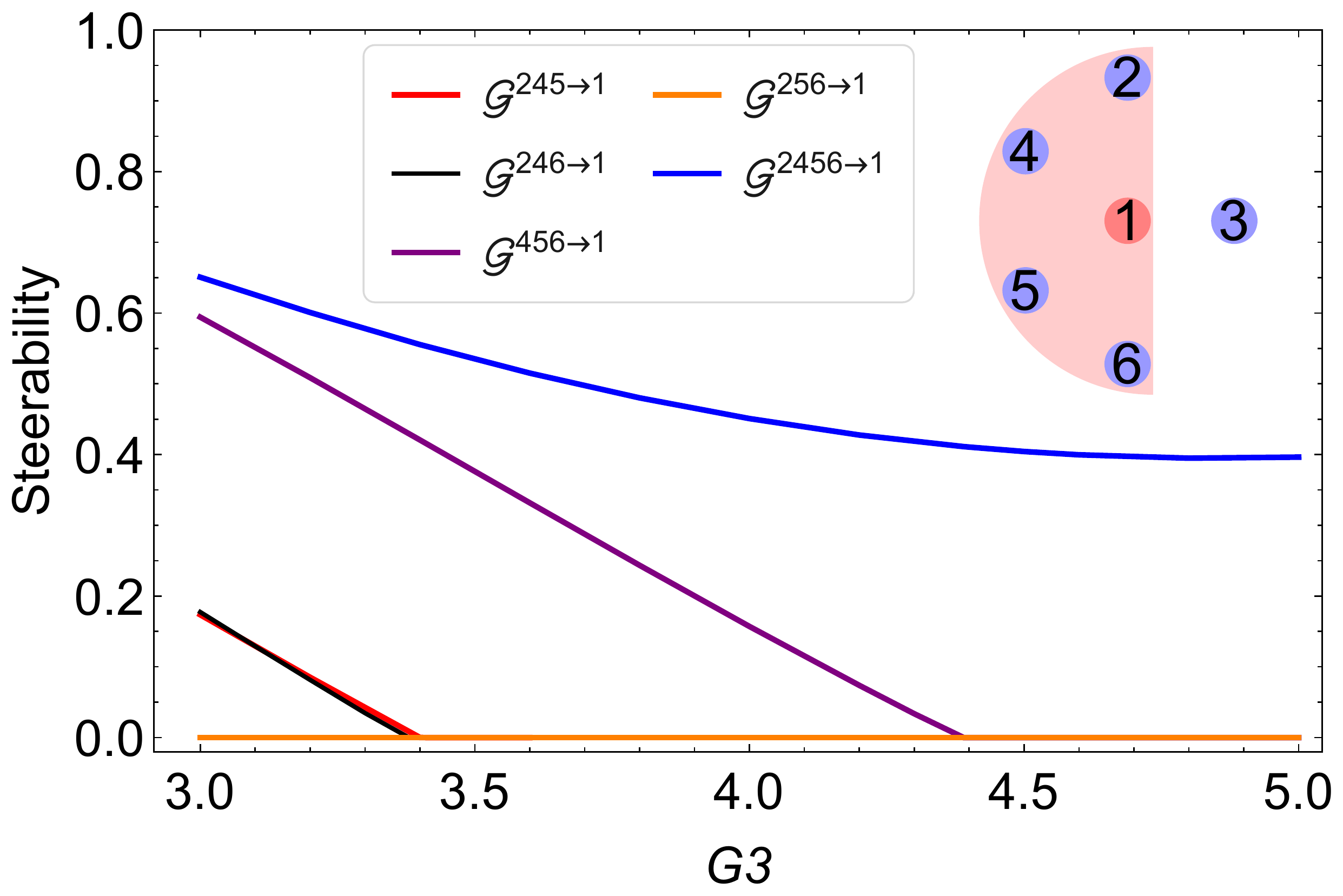}    
     \includegraphics[height=3.3cm,width=4.1cm]{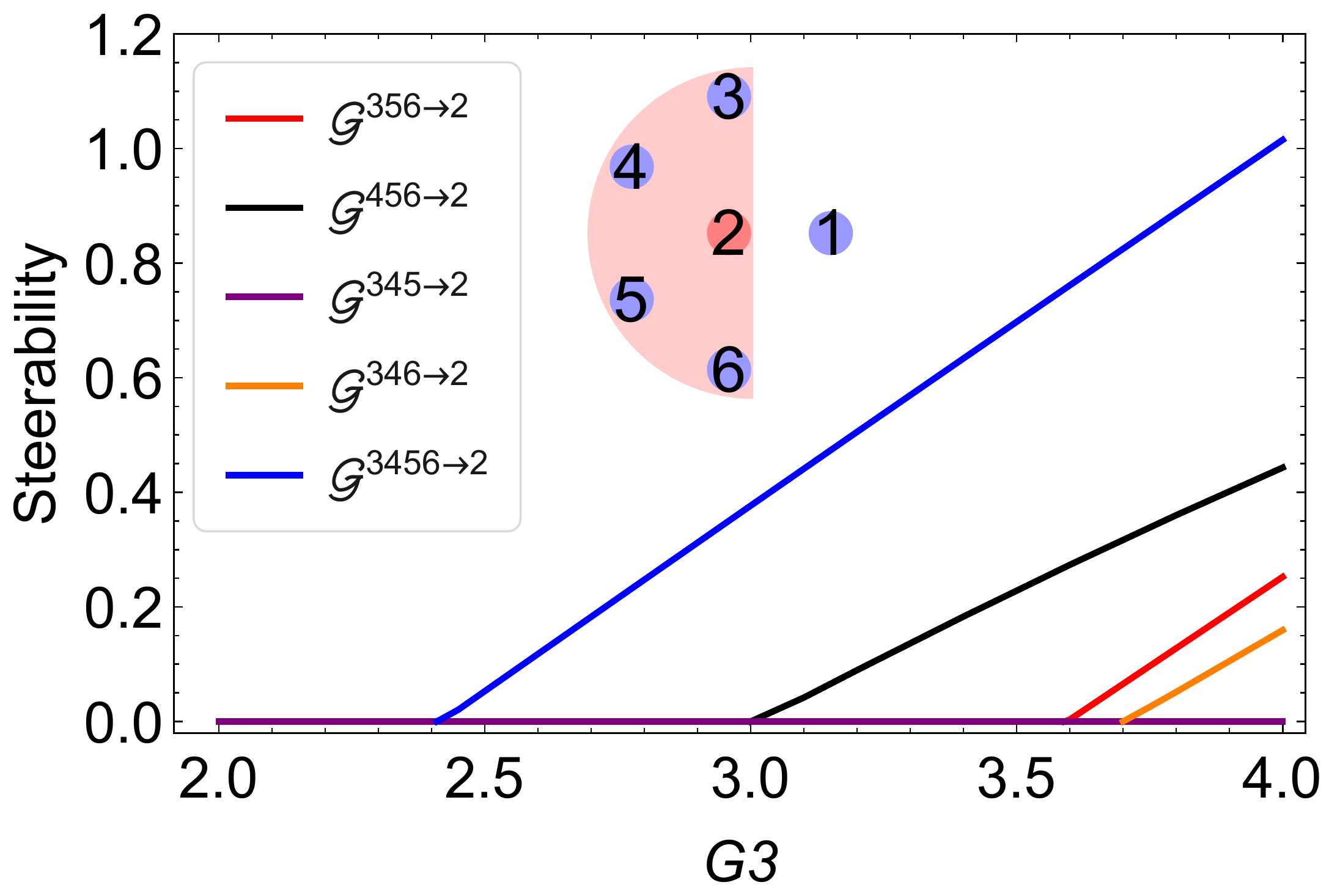}  
     \includegraphics[height=3.3cm,width=4.1cm]{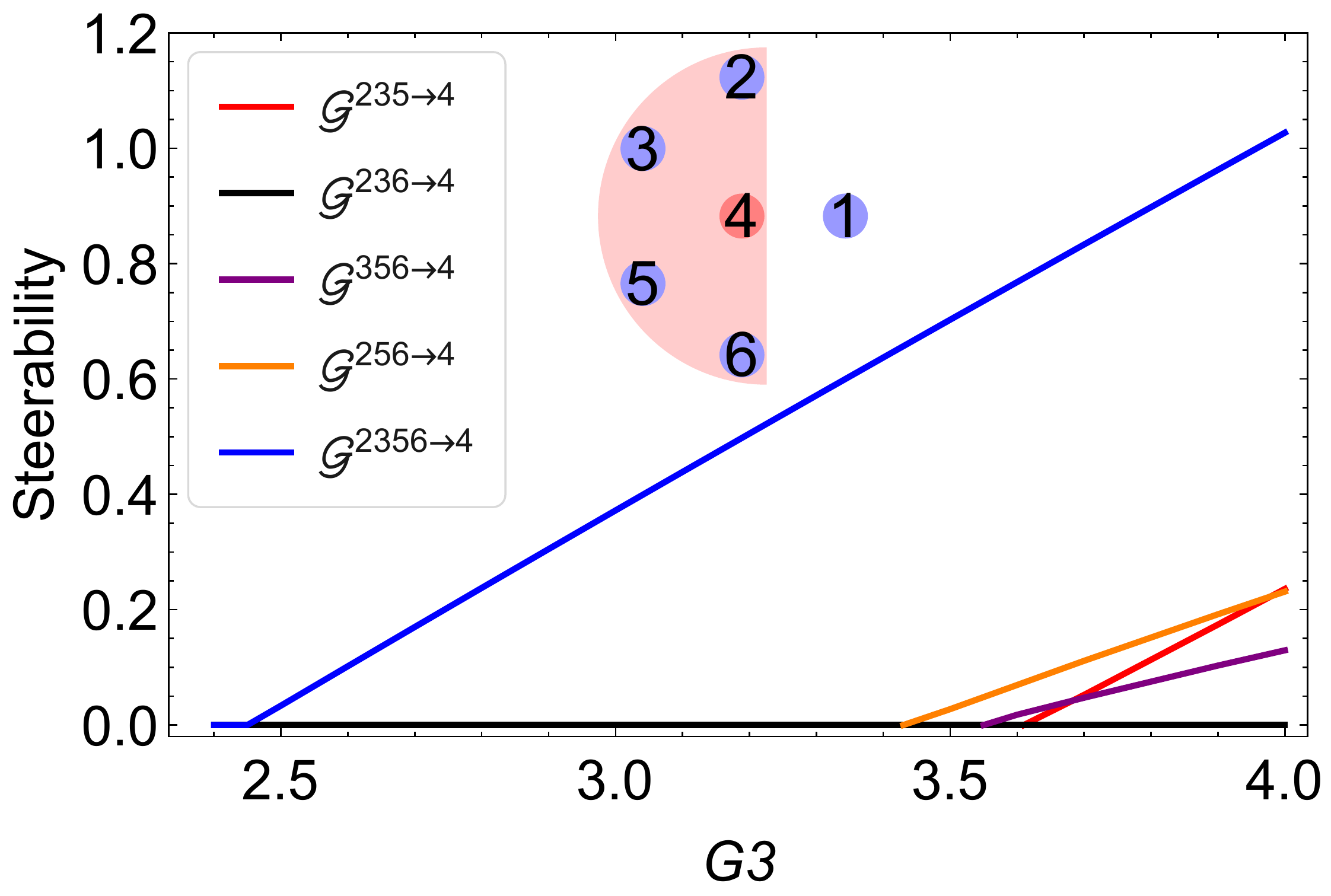} 
\\
\quad
\put(55,110){\bf($a^\prime$)}
\put(175,110){\bf($b^\prime$)}
\put(290,110){\bf($c^\prime$)}
\centering    
     \includegraphics[height=3.7cm,width=4cm]{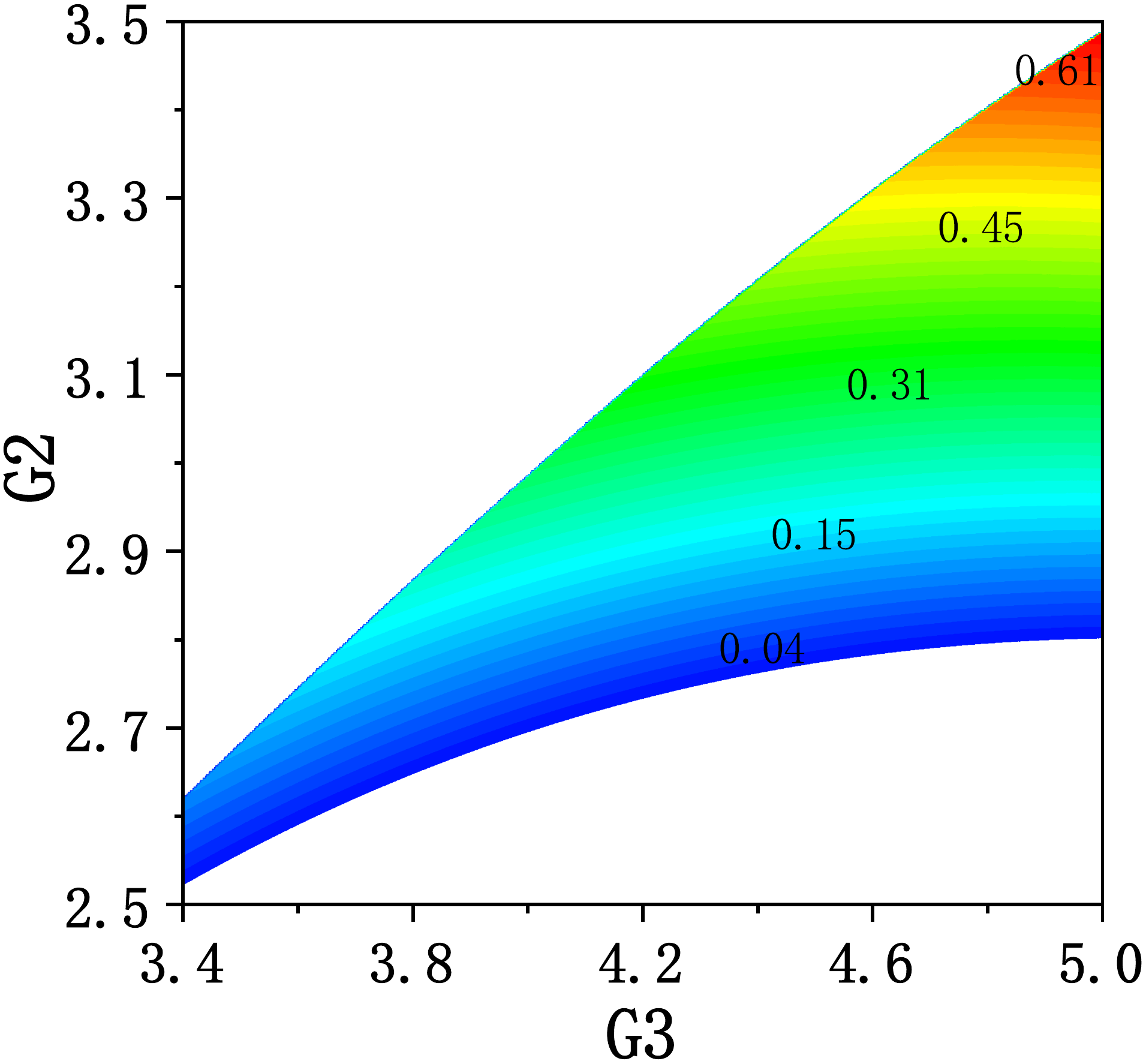}   
     \includegraphics[height=3.7cm,width=4cm]{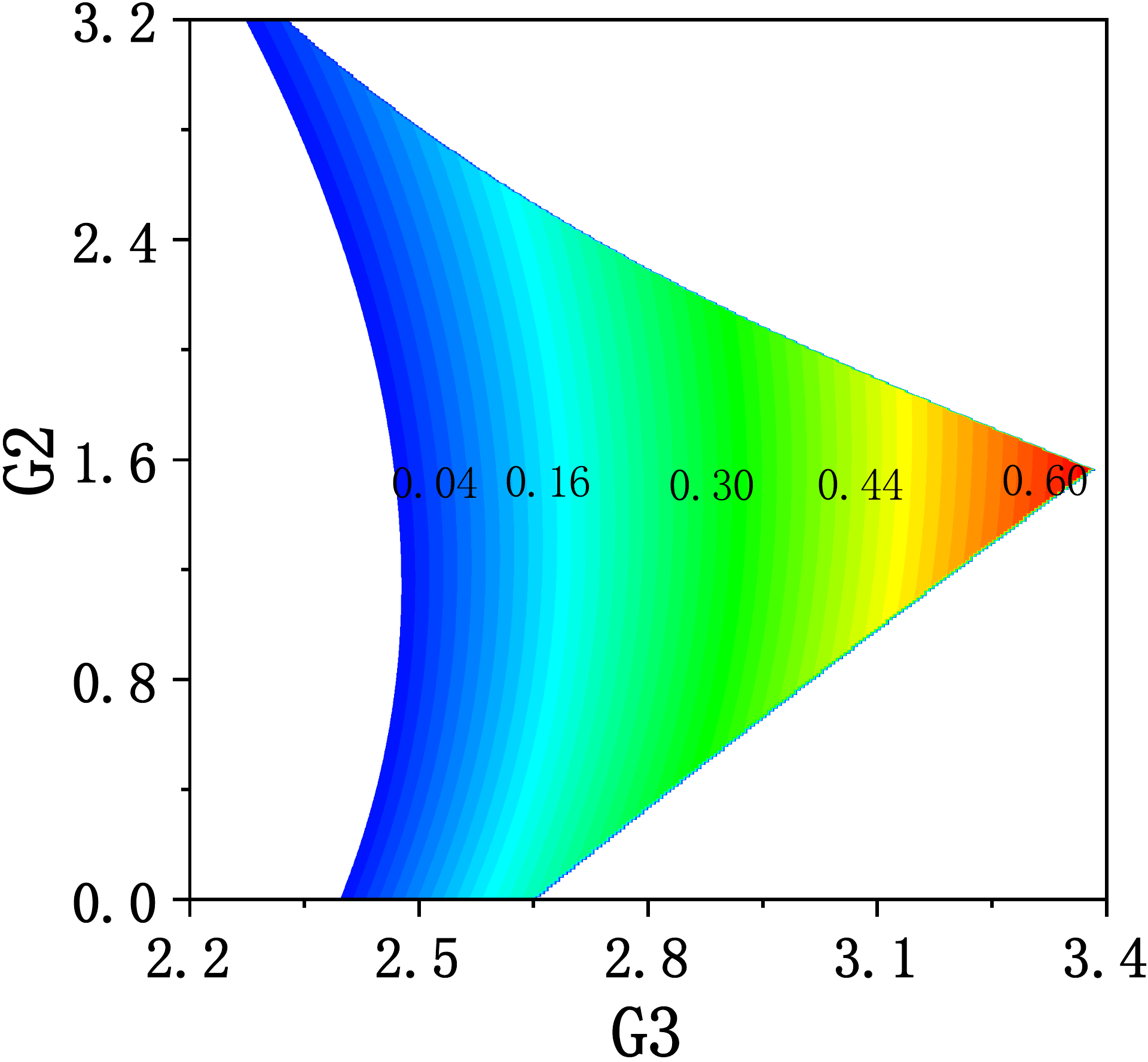}    
     \includegraphics[height=3.7cm,width=4cm]{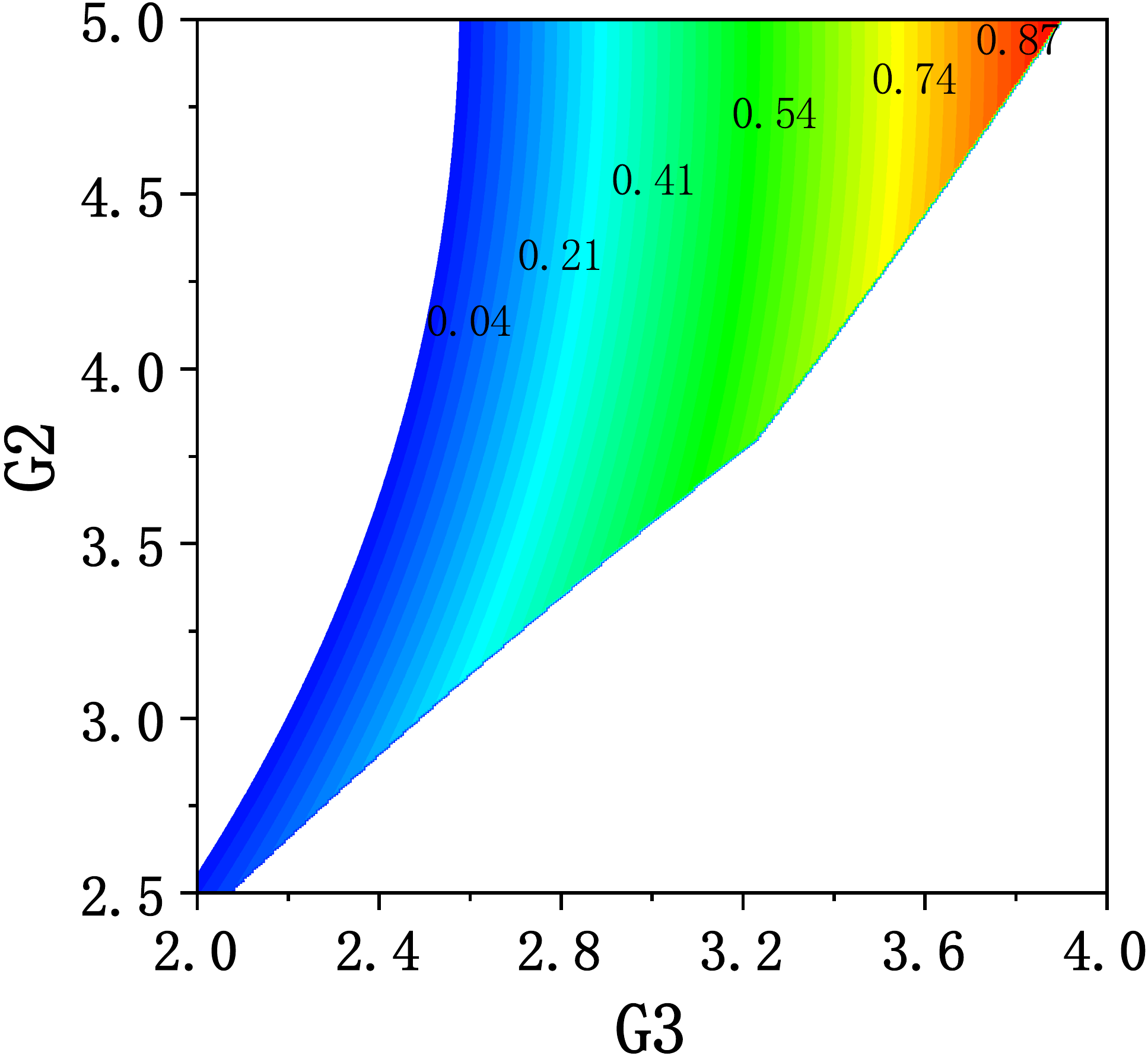}
\\  
  \centering    
     \includegraphics[height=3.7cm,width=4cm]{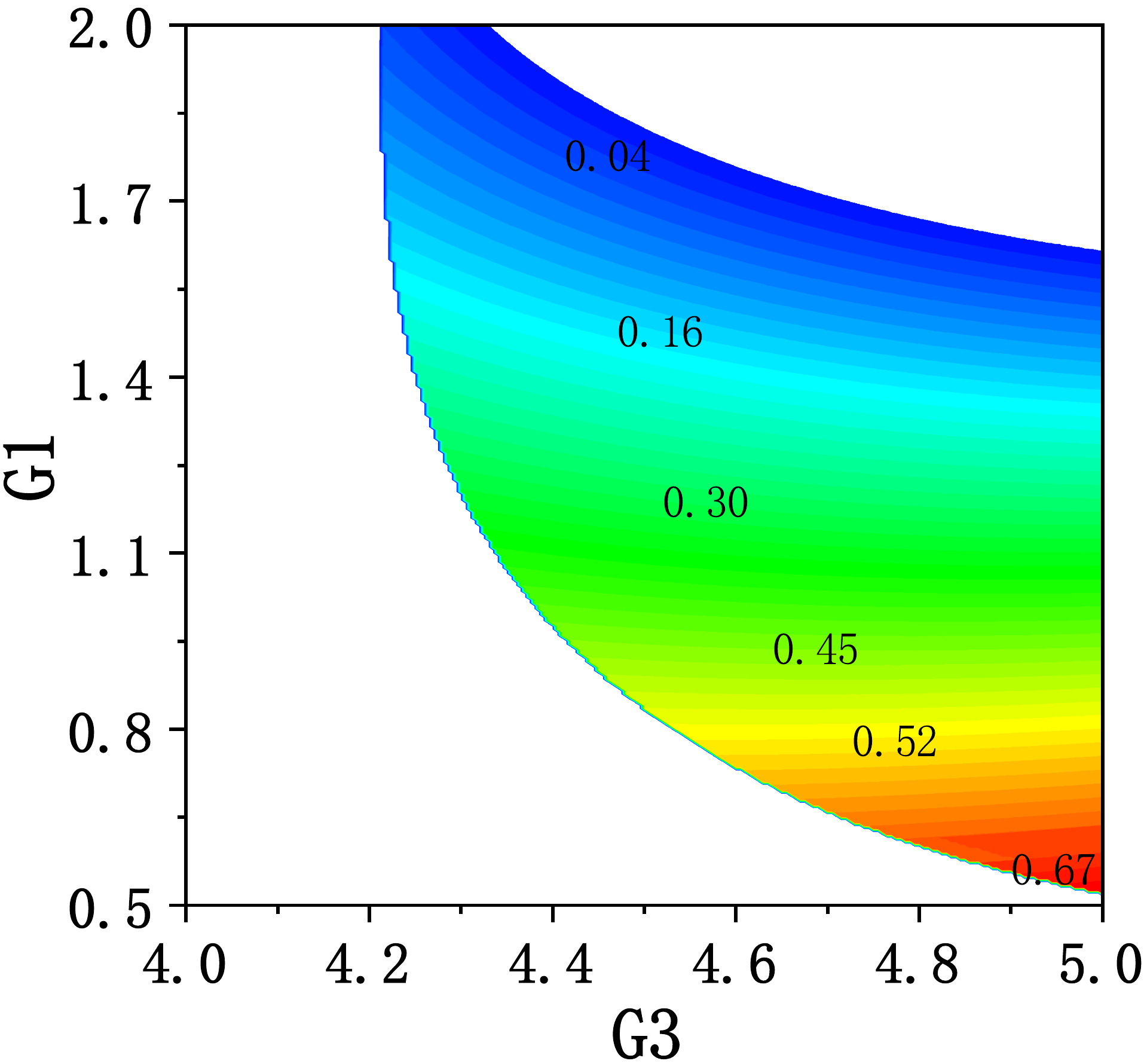}   
     \includegraphics[height=3.7cm,width=4cm]{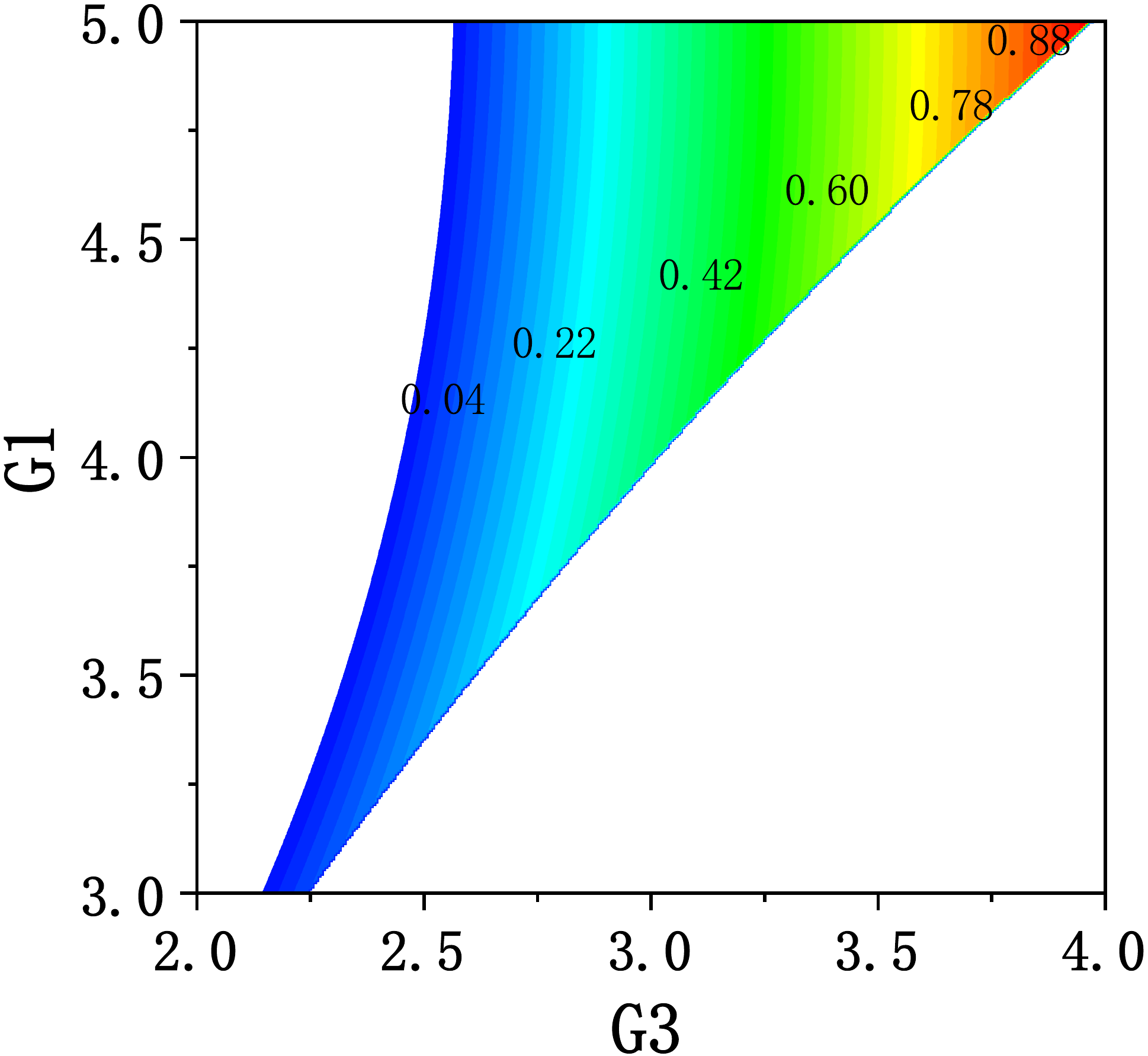}    
     \includegraphics[height=3.7cm,width=4cm]{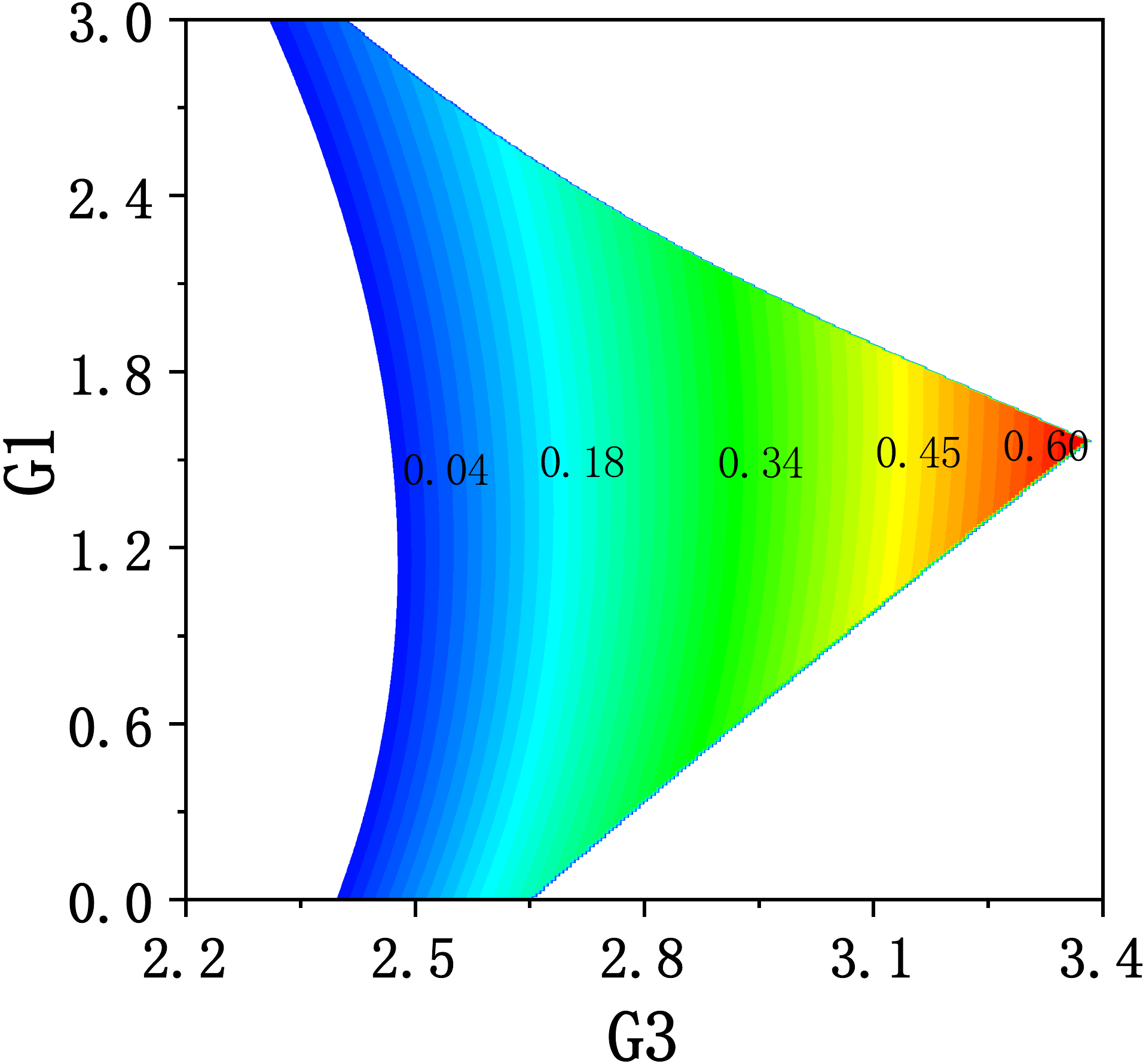}
  \quad
\put(-290,110){\bf($a^{\prime\prime}$)}
\put(-175,110){\bf($b^{\prime\prime}$)}
\put(-55,110){\bf($c^{\prime\prime}$)}   
\caption{ Parameter regions of the collective pentapartite steerings including modes $(\hat{a}_1 \hat{a}_2 \hat{a}_4 \hat{a}_5 \hat{a}_6)$ and $(\hat{a}_2 \hat{a}_3 \hat{a}_4 \hat{a}_5 \hat{a}_6)$. The steered mode in ($a$)/($a^\prime$)/($a^{\prime\prime}$), ($b$)/($b^\prime$)/($b^{\prime\prime}$), and ($c$)/($c^\prime$)/($c^{\prime\prime}$) are $\hat{a}_1$, $\hat{a}_2$, and $\hat{a}_4$, respectively. The corresponding parameters are: (a) $G_1=1,G_2=3.2$; (b) $G_1=4,G_2=2$; (c) $G_1=1.5,G_2=4$. ($a^\prime$)$G_1=1$,($a^{\prime\prime}$)$G_2=3.2$.($b^\prime$)$G_1=4$;($b^{\prime\prime}$)$G_2=2$.($c^\prime$)$G_1=1.5$; ($c^{\prime\prime}$)$G_2=4$.}
\end{figure}
The parameter range that the collective pentapartite steerings can exist are shown in Fig.6($a^{\prime}$)($b^{\prime}$)($c^{\prime}$) (collective pentapartite steering versus both $G_2$ and $G_3$ when $G_1$ is fixed) and Fig.6($a^{\prime\prime}$)($b^{\prime\prime}$)($c^{\prime\prime}$) (collective pentapartite steering versus both $G_1$ and $G_3$ when $G_2$ is fixed). It is worth mentioning that results of Fig.6(a)/(b)/(c) are included in Fig.6($a^{\prime}$)/($b^{\prime}$)/($c^{\prime}$) and Fig.6($a^{\prime\prime}$)/($b^{\prime\prime}$)/($c^{\prime\prime}$). For example, in Fig.6($a^{\prime}$), the parameter range of $G_2$ and $G_3$ of the collective pentapartite steering that only $\mathcal{G}^{2456\to 1}$ can happen, is given by the colorful zone, which includes the special condition shown in Fig6(a) ($G_1 =1,G_2 =3.2, G_3>4.3$). From six parameter regions, it is also noticed that steerabilities that larger than $\ln (e/2)$ (0.31) can be obtained under proper parameter conditions, therefore it can meets the condition of 1SDI QSS with nonzero key rates\cite{He2017QSS}. These results mean that one can choose five modes in the output six modes to build a collective pentapartite EPR steering, which includes more modes and have compacter setup, stronger steeribilities, than other works. 
\section{Monogamy relations}
Besides the collective steering, another key point of multipartite steering is the so-called monogomy relations, which is important for understanding how the steerings can be distributed among many parties and has been widely studied both theoretically and experimentally. There are four types of monogomy relations that have been developed\cite{Reid2013mo,He2020mo}, for Gaussian states with Gaussian measurements. Considering the six output modes in our model,  the four types of monogomy relations are shown in Table 2. As is mentioned in section 3.1, 3.2, and 3.3, all the type-I, type-II, and type-III monogomy relations can be satisfied in our model, which can be found in the corresponding matrix representations.
\begin{table}[h!]
\centering
\begin{center}
\caption{Monogamy Relations}
	\begin{tabular}{ | m{1cm}<{\centering}|m{5cm}<{\centering} | m{6cm}<{\centering} | m{4cm} |}
		\hline
  Type
    & Diagram &Condition \\ \hline	
		
\uppercase\expandafter{\romannumeral1}&\includegraphics[height=0.9cm,width=3.8cm]{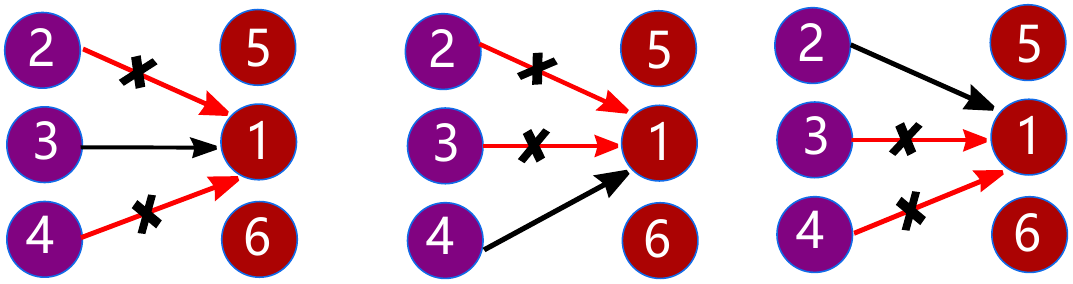} &$\mathcal{G} ^{3\to 1}>0,
\mathcal{G} ^{2\to 1}=0,\mathcal{G} ^{4\to 1}=0$\ $\mathcal{G} ^{4\to 1}>0,
\mathcal{G} ^{2\to 1}=0,\mathcal{G} ^{3\to 1}=0$\ $\mathcal{G} ^{2\to 1}>0,
\mathcal{G} ^{3\to 1}=0,\mathcal{G} ^{4\to 1}=0$ \\ \hline
		 \uppercase\expandafter{\romannumeral2}
    & \includegraphics[height=0.8cm,width=3.6cm]{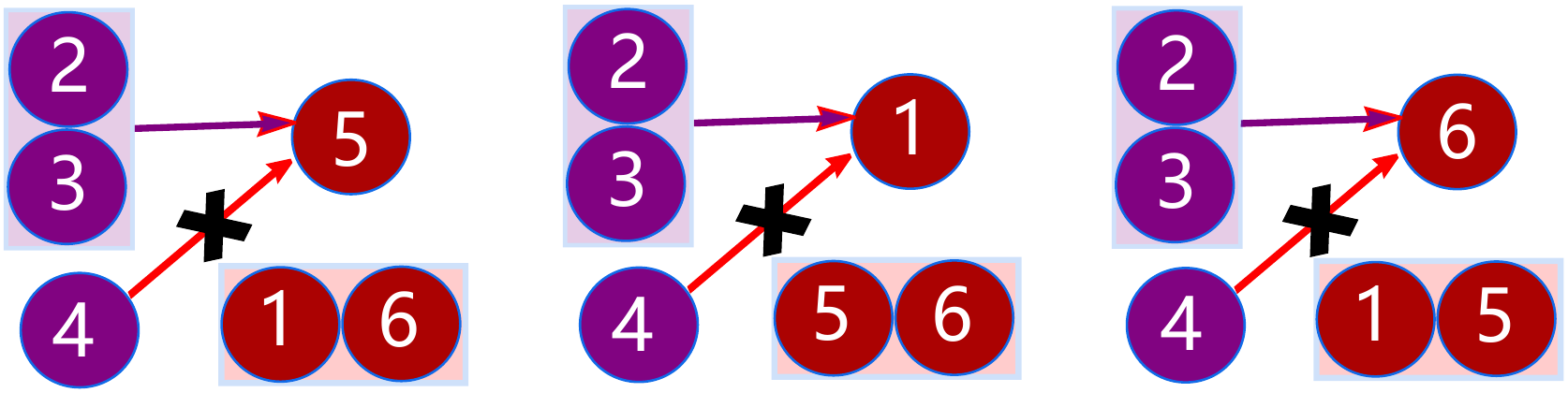} &$\mathcal{G} ^{23\to 5}>0,
\mathcal{G} ^{4\to 5}=0$      \        
   $\mathcal{G} ^{23\to 1}>0,
\mathcal{G} ^{4\to 1}=0$         \                 $\mathcal{G} ^{23\to 6}=0,\mathcal{G} ^{4\to 6}=0$ \\ \hline		
\uppercase\expandafter{\romannumeral3}a & \includegraphics[height=2.7cm,width=4.3cm]{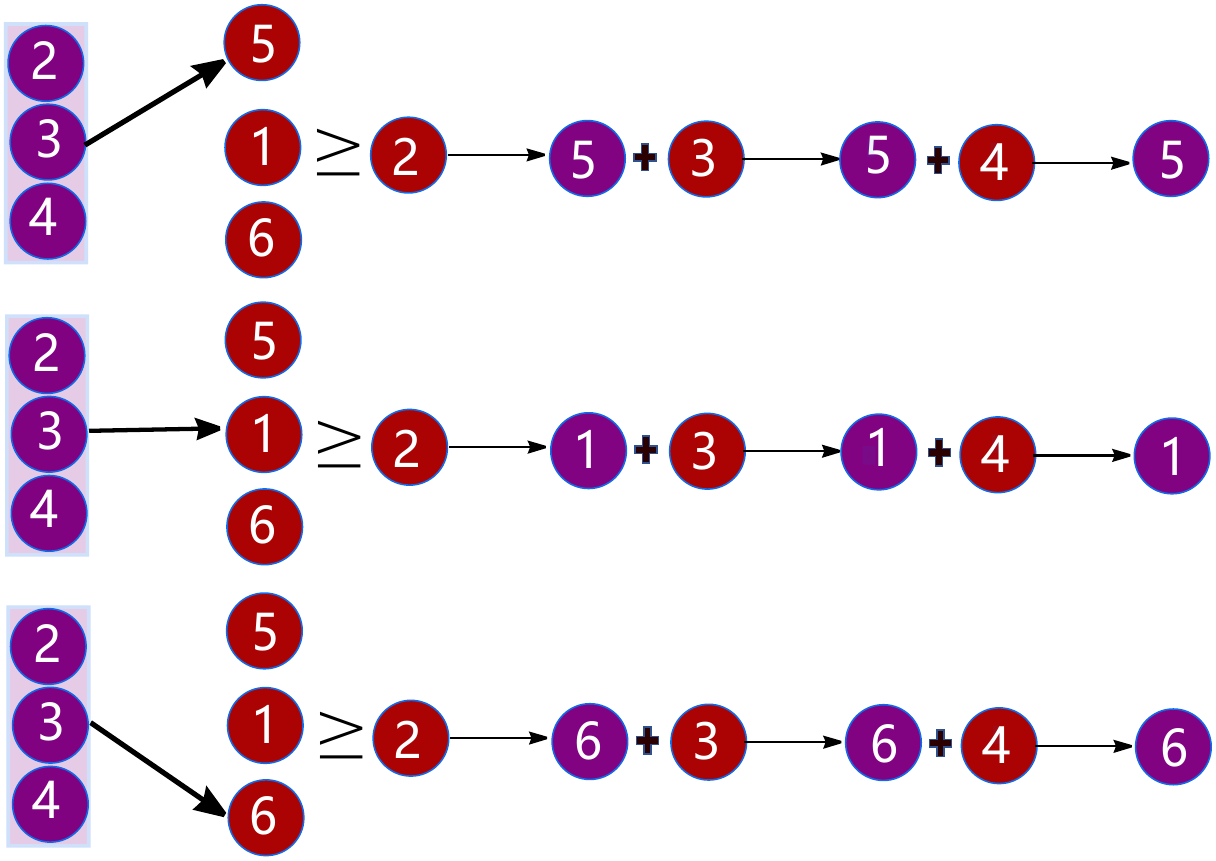} & $\mathcal{G} ^{234\to 5}-
\mathcal{G} ^{2\to 5}-\mathcal{G} ^{3\to 5}-\mathcal{G} ^{4\to 5}\geq0 $\ $\mathcal{G} ^{234\to 1}-
\mathcal{G} ^{2\to 1}-\mathcal{G} ^{3\to 1}-\mathcal{G} ^{4\to 1}\geq0 $\
$\mathcal{G} ^{234\to 6}-
\mathcal{G} ^{2\to 6}-\mathcal{G} ^{3\to 6}-\mathcal{G} ^{4\to 6}\geq0 $ \\ \hline	
\uppercase\expandafter{\romannumeral3}b & \includegraphics[height=2.7cm,width=4.3cm]{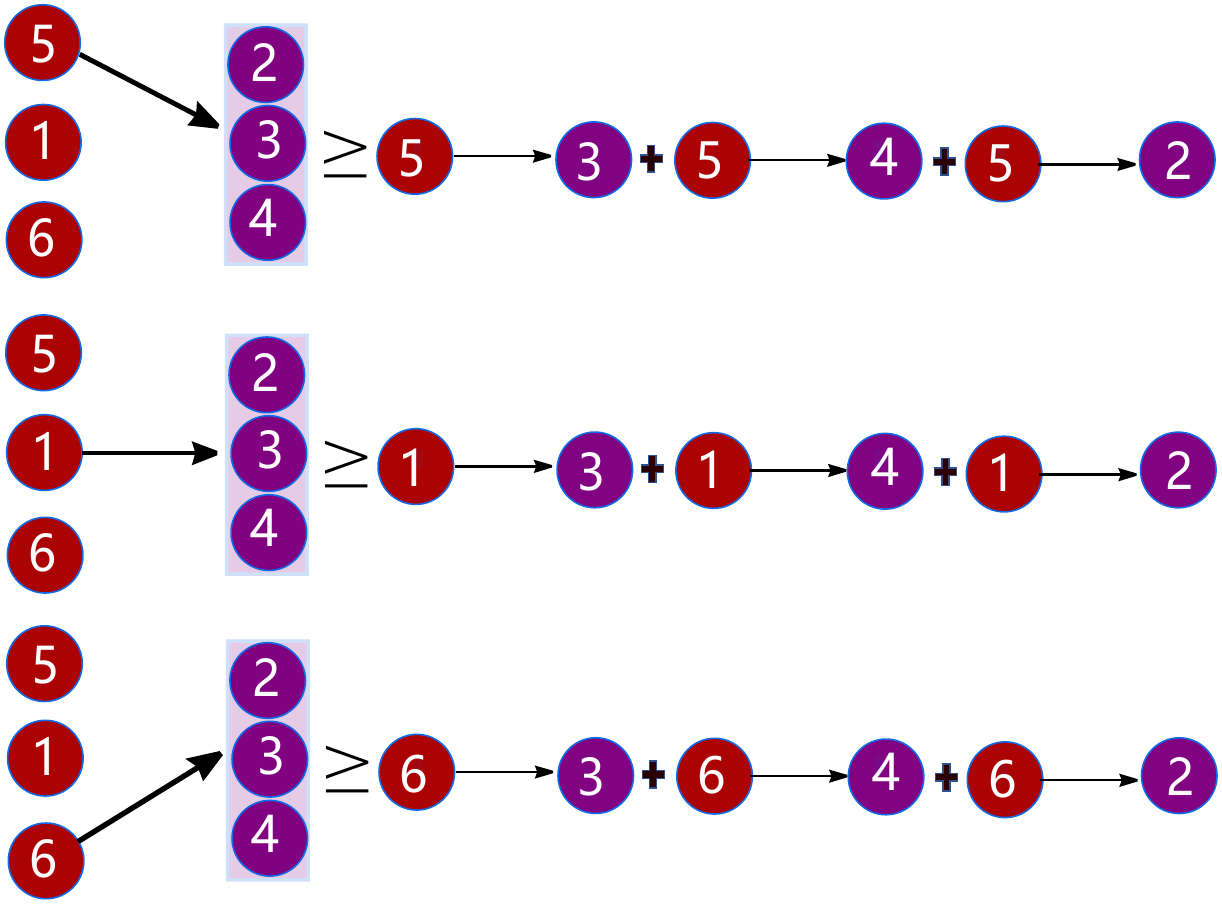} & $\mathcal{G} ^{5\to 234}-
\mathcal{G} ^{5\to 2}-\mathcal{G} ^{5\to 3}-\mathcal{G} ^{5\to 4}\geq0 $\ $\mathcal{G} ^{1\to 234}-
\mathcal{G} ^{1\to 2}-\mathcal{G} ^{1\to 3}-\mathcal{G} ^{1\to 4}\geq0 $\ 
$\mathcal{G} ^{6\to 234}-
\mathcal{G} ^{6\to 2}-\mathcal{G} ^{6\to 3}-\mathcal{G} ^{6\to 4}\geq0 $
\\ \hline	
\uppercase\expandafter{\romannumeral4}a
    & \includegraphics[height=0.8cm,width=4.4cm]{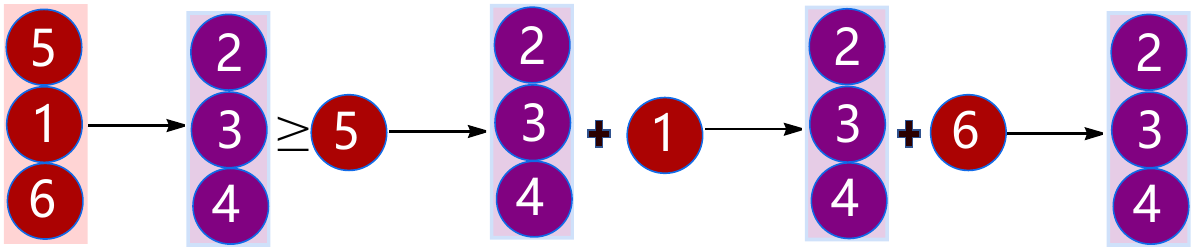}& $\mathcal{G} ^{156\to 234}-\mathcal{G} ^{5\to 234}-\mathcal{G} ^{1\to 234}-\mathcal{G}^{6\to234}\geq 0$ \\ \hline
    \uppercase\expandafter{\romannumeral4}b
    & \includegraphics[height=0.8cm,width=4.4cm]{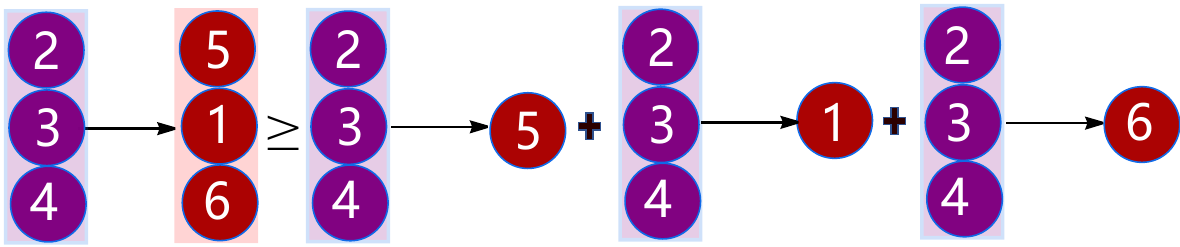}& $\mathcal{G} ^{234\to 156}-\mathcal{G} ^{234\to 1}-\mathcal{G} ^{234\to 5}-\mathcal{G}^{234\to6}\geq 0$ \\ \hline
	\end{tabular}
\end{center}
\end{table}
For our six-mode-system, the type-I and type-II monogomy relations mean that the steering parties cannot steer the steered party simultaneously, where the type-I monogomy relations correspond to the condition that all the steering and steered parties include one mode, and the type-II monogomy relations correspond to the condition that the steered party includes one mode and the steering parties include more than one modes, as are shown in the first and second row of Table 2. While the type-III monogomy relations mean that the steeribility from many modes to one mode is bigger than the sum of that from corresponding one mode to one mode, which can be found in the third row of Table 2. However, the type-IV monogomy relations is about steering from many modes to many modes, and steeribility from many modes to many modes is bigger than the sum of steeribilities from corresponding many modes to one mode, or from one mode to many modes. Since the type-IV monogomy relations have not been mentioned in the above sections, here we focus on them and mainly discuss about the situation that the steering part and steered part both have three modes, which is naturally included in our model. Considering a quadrapartition system ($n_A+n_B+n_C+n_D$)-mode system ABCD, the type-IV monogomy relation can be expressed as $\mathcal{G} ^{D \to (ABC)}-\mathcal{G} ^{D\to A}-\mathcal{G} ^{D\to B}-\mathcal{G}^{D\to C}\geq 0$, or $\mathcal{G} ^{(ABC) \to D}-\mathcal{G} ^{A\to D}-\mathcal{G} ^{B\to D}-\mathcal{G}^{C\to D}\geq 0$, with $ n_A=1, n_B=1, n_C=1; n_D=3 $ . The results of the type-IV monogomy relations, i.e., $\mathcal{G} ^{156\to 234}-\mathcal{G} ^{1\to 234}-\mathcal{G} ^{5\to 234}-\mathcal{G}^{6\to 234}$, and $\mathcal{G} ^{234\to 156}-\mathcal{G} ^{234\to 1}-\mathcal{G} ^{234\to 5}-\mathcal{G}^{234\to 6}$, versus parameters, are shown in Fig.7(a)(c) and Fig.7(b)(d), respectively. The colorful zones in Fig.7 correspond to the parameter range that the type-IV monogomy relations can be satisfied, i.e., the type-IV monogomy relations are true or not depend on the parameter conditions, which is different from other three types of monogomy relations. Therefore, for our model, the type-I, type-II, and type-III monogomy relations are definite. However, the type-IV monogomy relations are conditional, which can provide the possibility to control the distribution of the steering and may have potential applications in building controllable ultra-secure quantum network.

\begin{figure}[htbp]
\centering
\quad
\put(55,105){\bf($a$)}
\put(170,105){\bf($b$)}
  \centering    
     \includegraphics[height=3.7cm,width=4cm]{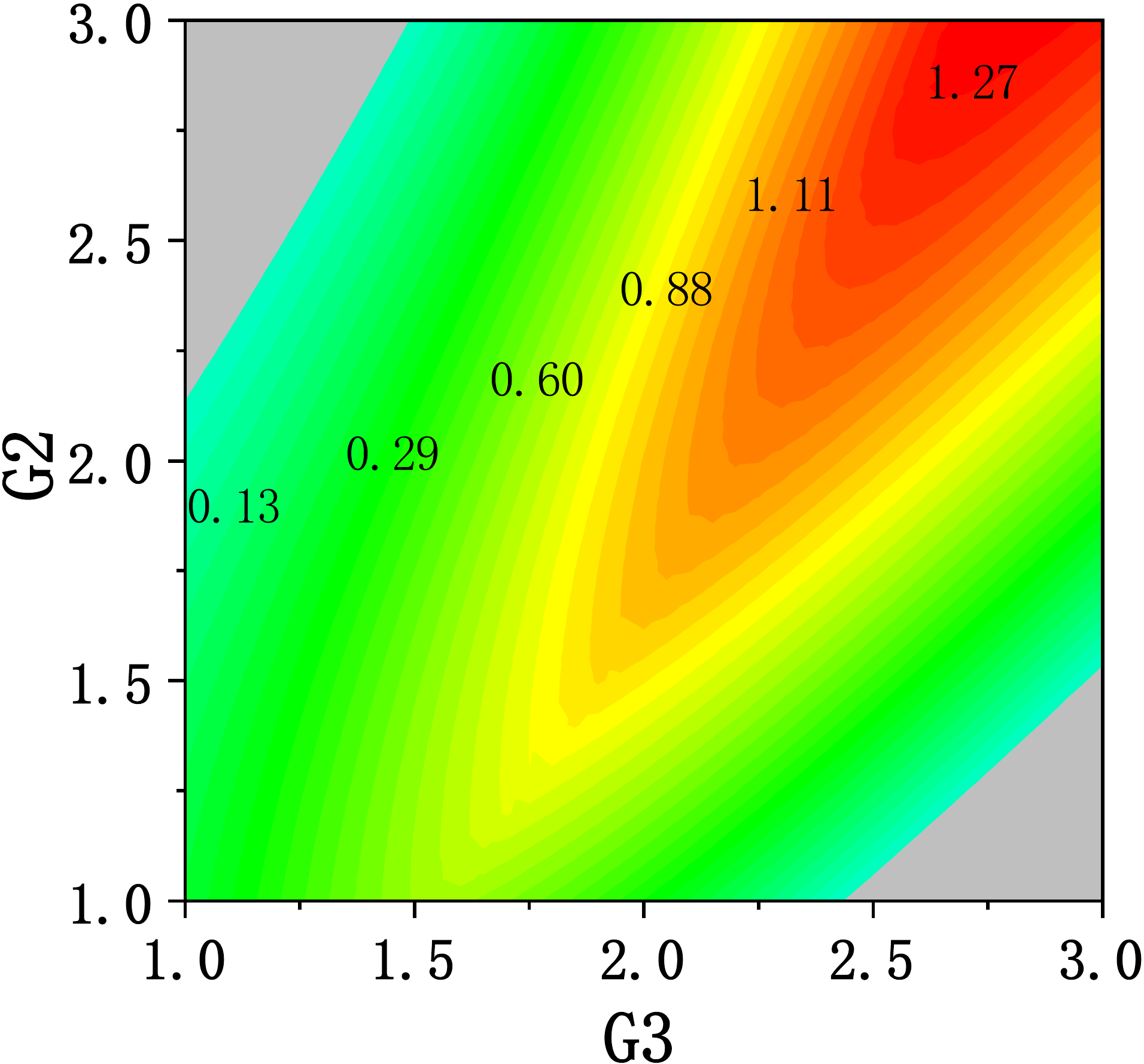} 
     \includegraphics[height=3.7cm,width=4cm]{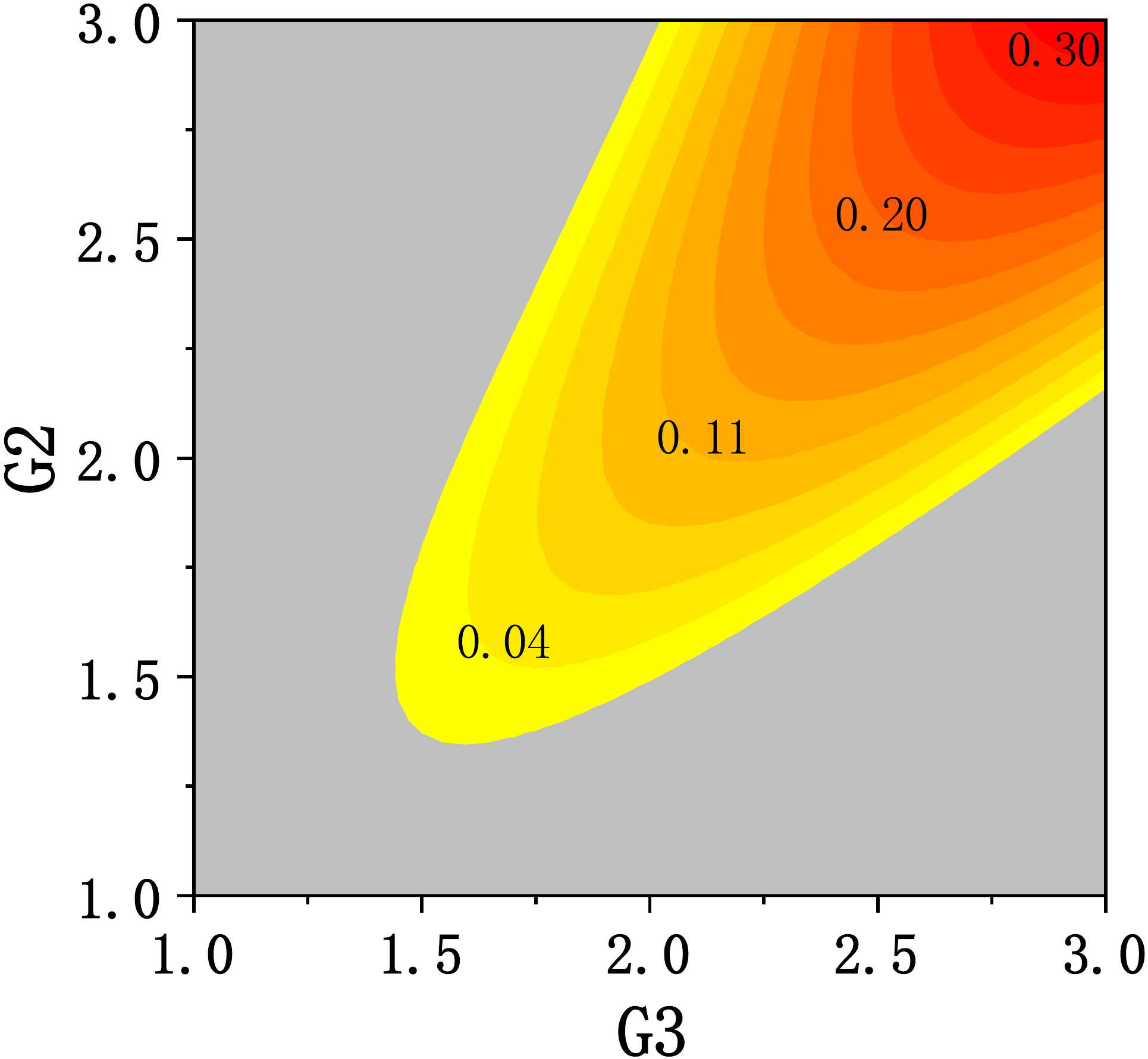} 
\\  
\quad 
\put(57,105){\bf($c$)}
\put(170,105){\bf($d$)}
\centering 
   \includegraphics[height=3.7cm,width=4cm]{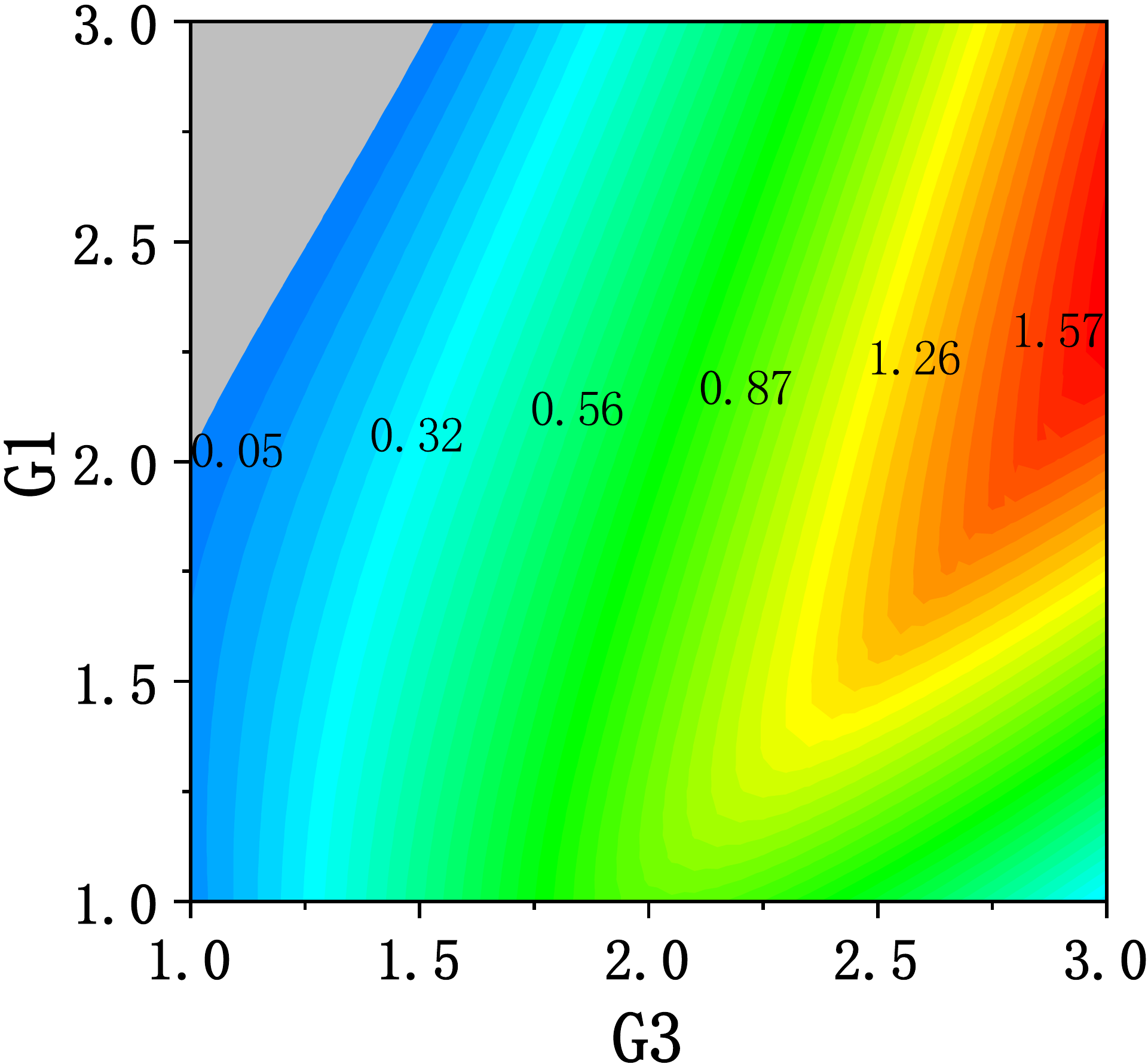}   
   \includegraphics[height=3.7cm,width=4cm]{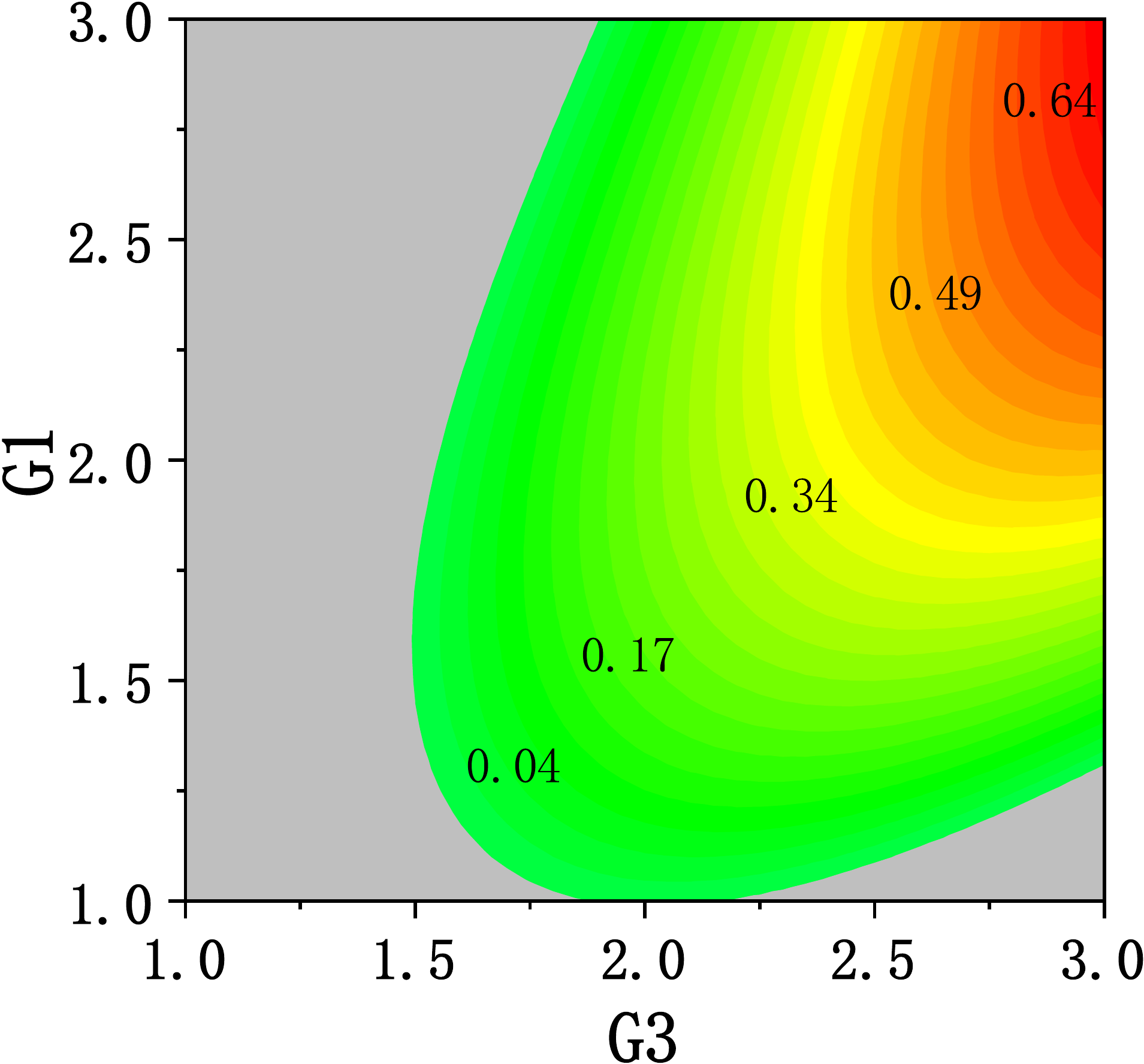}  
\caption{Parameter regions of type-IV monogomy relation, i.e., $\mathcal{G} ^{156\to 234}-\mathcal{G} ^{5\to 234}-\mathcal{G} ^{1\to 234}-\mathcal{G}^{6\to234}$ versus parameters for (a) $G_1=1.2$ and (c) $G_2=2$. $\mathcal{G} ^{234\to 156}-\mathcal{G} ^{234\to 1}-\mathcal{G} ^{234\to 5}-\mathcal{G}^{234\to6}$ versus parameters for (b) $G_1=1.2$ and (d) $G_2=2$}
\end{figure}
\section{Conclusion}
The steering properties of the six output beams that come from the FWM process with SSP have been studied in detail, including the (1+i)/(i+1)-mode steerings (i=1,2,3), the possible collective pentapertite steerings, and the interesting monogomy relations. It is found that the (1+i)/(i+1) steerings depend on the relative intensity of the related beams and the symmetry of the system, and bigger i will lead to less parameter dependence. Moreover, collective pentapartite EPR steerings can be obtained with a very compact setup, which are the central resource for ultra-secure hierarchical communication in quantum networks with many users where the issue of trust is of importance. In addtion, four types of the monogomy relations are discussed, and only the type-IV monogomy ralations are conditional, which provide the possibility to build controllable ultra-secure quantum network. All the results mean that this FWM process with SSP is a promising platform to demonstrate different kinds of EPR steerings and to explore the corresponding valuable applications. By the way, matrix representation is used to express the steerings for the first time, which is not only help to demonstrate all the steerings intuitively, but also very useful to understand the monogomy relations.


\bibliography{sample}






\end{document}